\documentclass[10pt,aps,amsmath,amssymb,prd,twocolumn,preprintnumbers,superscriptaddress]{revtex4-2}

\usepackage{orcidlink}

\begin{document}
\title{Correlated and coincident noise in Einstein Telescope's triangular configuration}

\author{Jan Harms\,\orcidlink{0000-0002-7332-9806}}
\email{jan.harms@gssi.it}
\affiliation{Gran Sasso Science Institute (GSSI), I-67100 L'Aquila, Italy}
\affiliation{INFN, Laboratori Nazionali del Gran Sasso, I-67100 Assergi, Italy}

\begin{abstract}
The Einstein Telescope (ET) is a proposed underground infrastructure in Europe to host future generations of gravitational-wave (GW) detectors. Two configurations are under investigation. The first is the two-site solution with an L-shape detector at each site similar to the LIGO configuration. The second configuration is the single-site triangular detector. While numerous analyses over the last decade have already proven that the distributed network of GW detectors is the scientifically preferred solution, it was also argued that the advantage might not be decisive enough to dismiss ET's triangle configuration. The modeling of aspects such as noise correlations in these configurations has only been treated superficially in the past. In this article, we introduce a time-domain simulation of environmental noises in the ET triangle using available data from ET's candidate sites. We simulate and analyze correlated environmental noise as well as coincident glitches produced by transients in the environmental fields. We find that these features pose an important limitation to the observational capabilities of the ET triangle and limit the utility of its null stream.
\end{abstract}

\maketitle
\section{Introduction}
The configuration of a gravitational-wave (GW) detector is a critical choice that needs to be made early in the project development because of its profound impact on a detector's observational capabilities, infrastructure, and operation. The standard concept for a laser-interferometric GW detector is the configuration with two arms at a right angle, which first appeared in a publication in 1971 \cite{Moss1971}; henceforth called the L-shape configuration. The two arms serve to suppress (practically cancel) noise in the GW measurement coming from the laser by destructive optical interference towards the measurement port, while the right angle maximizes the detector response to the quadrupolar GW polarization. This configuration was chosen for today's ground-based detectors LIGO, Virgo, and KAGRA (LVK) \cite{Capote2025,AdvVirgo2015,KAGRA2021}. 

Already before the LVK projects came into reality, configurations with angles different from 90$^\circ$ had been proposed. The L-shape configuration turned out to be a bad solution for space-based detectors, because an L-shape formation of satellites in a solar orbit can be maintained only at a great expense of fuel. \citet{Faller1985} proposed a 120$^\circ$ angle between the arms to solve this problem. In the same year, a study by \citet{Winkler1985} came out proposing the equilateral triangle configuration for a GW detector in Germany consisting of 3 interferometers. The arguments in favor of this configuration were redundancy to increase the overall observation time with at least two interferometers online, and linear dependence of GW signals (effectively the argument behind the ET null stream; see below). The triangle configuration with laser links between all three vertices soon became the winning configuration of the space-based detector LISA \cite{LISA2024}. It added redundancy to the Faller et al concept --- which is especially attractive to space concepts --- without sacrificing sensitivity to GWs.

While redundancy is not a major concern in ground-based detectors, two features make the triangle configuration interesting to GW science. Compared to a single L-shape detector, the triangle improves the observational capabilities by enabling the simultaneous measurement of both GW polarizations provided that all three laser interferometers are functional \cite{Cutler1998}. Furthermore, the GW signal is strongly suppressed (virtually absent) in the sum of signals of all three interferometers at frequencies $f\ll c/L$, where $L$ is the length of the interferometer arm. This sum, which is known as the null stream, was first analyzed in the context of GW detector networks \cite{GuTi1989}, and later found in the triangle configuration as so-called fully symmetric Sagnac and T-channel of the LISA detector \cite{TAE2000,PrEA2002}. In 2010, the ET concept was published and with the goal in mind to create the scientifically most potent concept for a detector, the triangular configuration was proposed \cite{Punturo2010,Freise2011}. 

In the meantime, the situation in Europe has evolved with growing support for a 2-site solution with detectors in the L-shape configuration. Scientifically, a triangle is superior to a single L-shape detector, but not to two L-shape detectors \cite{Branchesi2023a,Abac2026}. In fact, the triangle is equivalent to two collocated L-shape interferometers with 45$^\circ$ relative orientation (see section \ref{sec:nullspace}). The triangle might require complex scheduling of maintenance and upgrades to cope with personnel limitations and to avoid interference between site activities and interferometers being in observing runs. Moreover, the triangle bears the risk of having to sacrifice the independence of noise. The consequences of collocating GW detectors has been studied extensively in the case of the two LIGO Hanford interferometers \cite{Fotopoulos2008}. Noise correlations were observed between the two interferometers at Hanford, but no method was effective to reduce the noise correlations resulting in a fundamental sensitivity limitation of stochastic GW searches when using data from the two Hanford interferometers. Moving forward with both detectors was eventually realized to constitute a substantial risk to the LIGO project, which led to the decision to find a partner country that would be interested in using the components of the second Hanford detector for a new GW detector. This effort has become the LIGO India project \cite{Unni2024}. Consequently, it is crucial to analyze the complex dependencies created in the ET triangle with respect to noise and operations.

In this paper, we revisit the issue of correlated and coincident noise in the ET triangle and present results from a new time-domain simulation of ET triangle data. The ET detector is proposed as a xylophone composed of a low-frequency (ET-LF) and a high-frequency (ET-HF) interferometer; the full triangle thereby consists of 6 interferometers. Correlated and coincident noise caused by the environment can be expected to occur in both, ET-LF and ET-HF, but the problem is far more severe for ET-LF \cite{AmEA2020}. In section \ref{sec:nullspace}, we briefly recap the main properties of the ET triangle and the definition of its null stream. We present models of environmental-noise correlations in section \ref{sec:correl} and how to simulate time series with noise based on a cross spectral-density matrix. The impact of noise correlations on the signal-to-noise ratio of GW signals was first discussed in \cite{wong2024}, and we will revisit it in section \ref{sec:corrimpact} with focus on environmental noise. As part of the site-characterization campaigns, data from environmental fields including seismic and magnetometer data are recorded. Using environmental data for the noise models, we automatically obtain a simulation of transient instrument noise, as we show in section \ref{sec:transient}. Finally, in section \ref{sec:null}, we reanalyze the utilization of the null stream for the mitigation of noise transients as was proposed in \cite{GNH2022,Narola2025}.

\section{The ET null stream: a fundamental signal redundancy of the triangle configuration}
\label{sec:nullspace}
\begin{figure}[ht!]
    \centering
    \includegraphics[width=0.8\columnwidth]{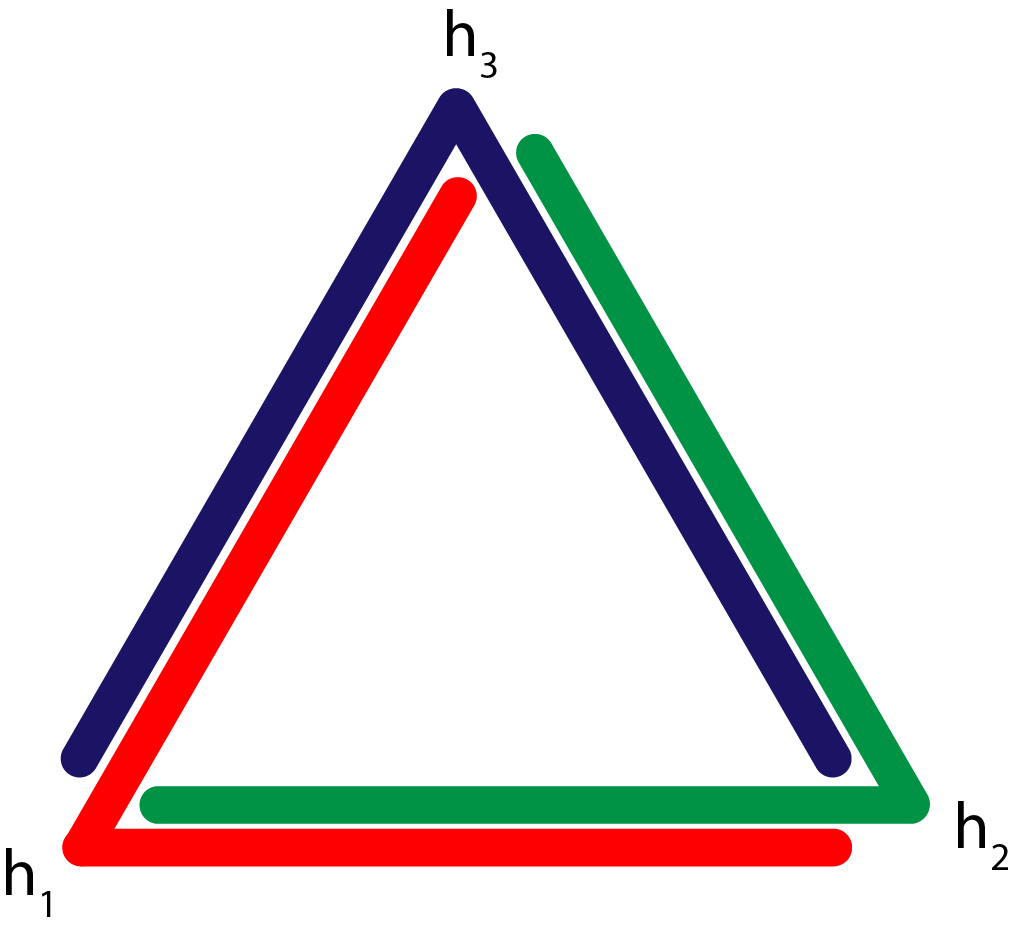}
    \caption{Sketch of the $\Delta$-configuration of the Einstein Telescope.}
    \label{fig:etconf}
\end{figure}

The Einstein Telescope was originally conceived in a triangular configuration  (represented by the symbol $\Delta$) as shown in figure \ref{fig:etconf}. Laser interferometric systems are centered at the three vertices of an equilateral triangle. Each system consists of two long-baseline interferometers in the so-called xylophone configuration \cite{HiEA2009}.
One interferometer is optimized for low-frequency operation subsequently called ET-LF, the other interferometer is optimized for high-frequency operation subsequently called ET-HF. In total, the  triangle configuration consists of 6 long-baseline laser interferometers. Each V-shape in figure \ref{fig:etconf} represents an ET-LF/HF pair.

Just by looking at the laser-interferometric combination, it should be clear that the GW signals observed with each interferometer are not independent. In fact, in lowest order of $L/\lambda$, where $\lambda$ is the length of the GW and $L$ is the length of an interferometer arm, the sum of the three signals vanishes:
\begin{equation}
    \sum\limits_{k=1}^3h_k=0.
    \label{eq:nullstream}
\end{equation}
We refer to \cite{Sch1997,EVE2017,Virtuoso2025} for response studies beyond the long-wavelength approximation $\lambda\gg L$. The authors showed that towards higher frequencies a propagation-direction and polarization independent null stream of the triangle becomes an increasingly poor approximation. For the purpose of this article, we assume that GW signals are virtually absent in the null stream. This reduces the signal space from 3 to 2 dimensions \cite{PrEA2002,GNH2022,WoLi2022}, which means that the triangle effectively corresponds to two laser-interferometric measurements of GW signals assuming for now that the noise in each measurement is independent of the others.

\begin{figure}[ht!]
    \centering
    \includegraphics[width=\columnwidth]{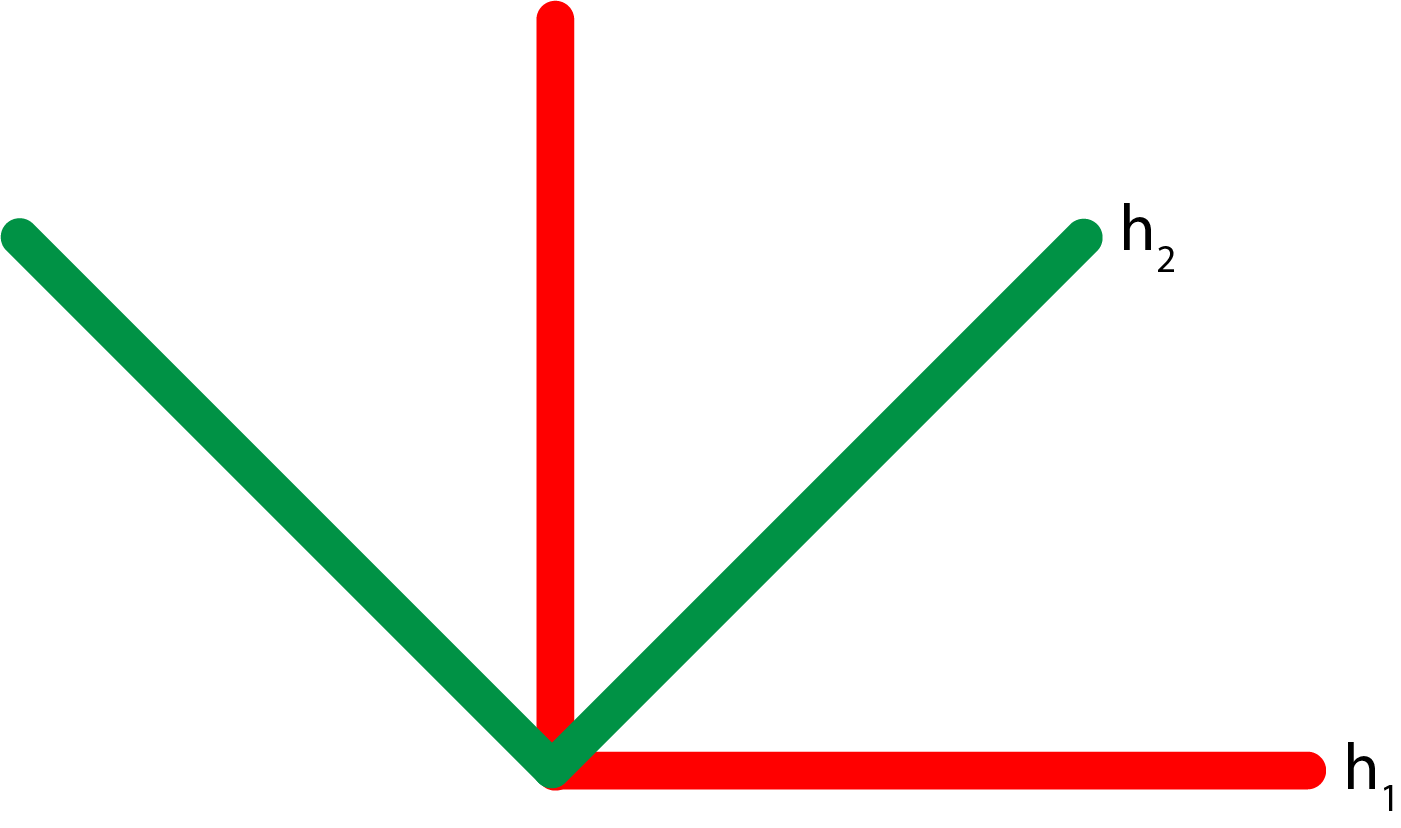}
    \caption{Effective 2L-representation of the ET triangle configuration.}
    \label{fig:etequiv}
\end{figure}

If the signal space of the triangle configuration has 2 dimensions, then this means that effectively two laser-interferometric measurements are done. The question is what type of laser interferometers these two measurements represent. It turns out that the signal space of the triangle configuration is equivalent to the signal space of two collocated L-shape detectors at 45$^\circ$ as shown in figure \ref{fig:etequiv}.
In other words, two such L-shape detectors retrieve exactly the same information about the GW as the triangle configuration. To be more precise, if the two L-shape detectors had the same length $L$ and the same strain sensitivity as the interferometers of the $\Delta$-configuration, then the total signal-to-noise ratio (SNR) of the two L-shape detectors would be smaller by a factor $2\sqrt{2}/3\approx 0.94$ compared to the $\Delta$-configuration:
\begin{equation}
    \sum\limits_{k=1}^3{\rm SNR}^2_k\bigg|_\Delta=\frac{9}{8}\sum\limits_{k=1}^2{\rm SNR}^2_k\bigg|_{\rm 2L,45^\circ}.
\end{equation}
The derivation of this result is straight-forward and simply requires the projection of the GW tensor $\mathbf h$ using the equation
\begin{equation}
    h_k=(\vec e_1^{\,k}\otimes\vec e_1^{\,k}-\vec e_2^{\,k}\otimes\vec e_2^{\,k}):\mathbf h,
\end{equation}
where $\vec e_1^{\,k},\,\vec e_2^{\,k}$ are the unit vectors pointing along the two arms of detector $k$. In our case, the same tensor $\mathbf h$ is used to calculate the signal of each detector since all detectors are collocated. The squares of the signals $h_k$ need to be added for all 2 or 3 detectors to calculate the total SNR. The SNR ratio holds for each individual GW signal.

Note that the 2L-configuration of ET was proposed with arm lengths of 15\,km, which means that the 2L-configuration has an SNR advantage over the 10\,km triangle in addition to the profit that comes from the distance between the two detectors.

\section{Environmental correlations}
\label{sec:correl}
The goal of this section is to first provide some understanding of the physically possible values of coherence between interferometers of the $\Delta$-configuration if caused by environmental fields. Operation of the current Virgo and LIGO detectors has shown that environmental fields can couple with the detector through a plethora of mechanisms \cite{EfEA2015,FiEA2020}. The main impact is on the detector sensitivity in the low-frequency band. Environmental disturbances can cause noise correlations in detector networks and interfere with GW analyses \cite{AaEA2015,TCS2013,CoEA2016b,AnHa2020,JaEA2022,Branchesi2023a,JaEA2024,BHR2024}. 

To explore physically realistic coherence values between detectors of the $\Delta$-configuration, we will go through a few  simple examples, which highlight the key aspects of the problem. The signal measured by each interferometer can be written 
\begin{equation}
    s_k=(x_k^{22}-x_k^{21})-(x_k^{12}-x_k^{11})+n_k,
    \label{eq:diff}
\end{equation}
where $k=1,2,3$ is the detector index, and $x_k^{ij}$ is the displacement of mirror $j=1,2$ in arm $i=1,2$, and $n_k$ is readout noise such as shot noise. For the purpose of this section, we can neglect the contribution of the readout noise ($n_k=0$), and all displacement noises $x_k^{ij}$ are assumed to have the same power spectral density (PSD). The coherence between displacement noises will be denoted as $\langle x_1^{12}|x_2^{21}\rangle$ with $0<|\langle x_1^{12}|x_2^{21}\rangle|<1$ and similarly for other test-mass pairs. The examples are presented in table \ref{tab:correl}. 

We present the calculation of scenario 1 in detail, and all other scenarios can be evaluated analogously. We start by correlating data of two ET-LF interferometers: 
\begin{equation}
  \langle s_1|s_2\rangle = \langle x_1^{11}|x_2^{22}\rangle=-\langle F_1^{11}|F_2^{22}\rangle=-1
\end{equation}
All other contributions to the correlation vanish in this scenario. Note that the minus sign comes from the fact that the axes of the coordinate system, in which the displacements of test masses of the two interferometers are evaluated, point in opposite directions, while the environmental forces at the two test masses are evaluated in a common coordinate system. In order to obtain the coherence, we also need the noise PSDs,
\begin{equation}
    \langle s_1|s_1\rangle =  \langle s_2|s_2\rangle =4 \langle x_1^{11}|x_1^{11}\rangle=4 \langle x_2^{22}|x_2^{22}\rangle = 4,
\end{equation}
and the coherence is given by
\begin{equation}
    \gamma_{12} = \frac{\langle s_1|s_2\rangle}{\sqrt{\langle s_1|s_1\rangle\langle s_2|s_2\rangle}}=-0.25
\end{equation}

\begin{figure}
    \centering
    \includegraphics[width=0.9\columnwidth]{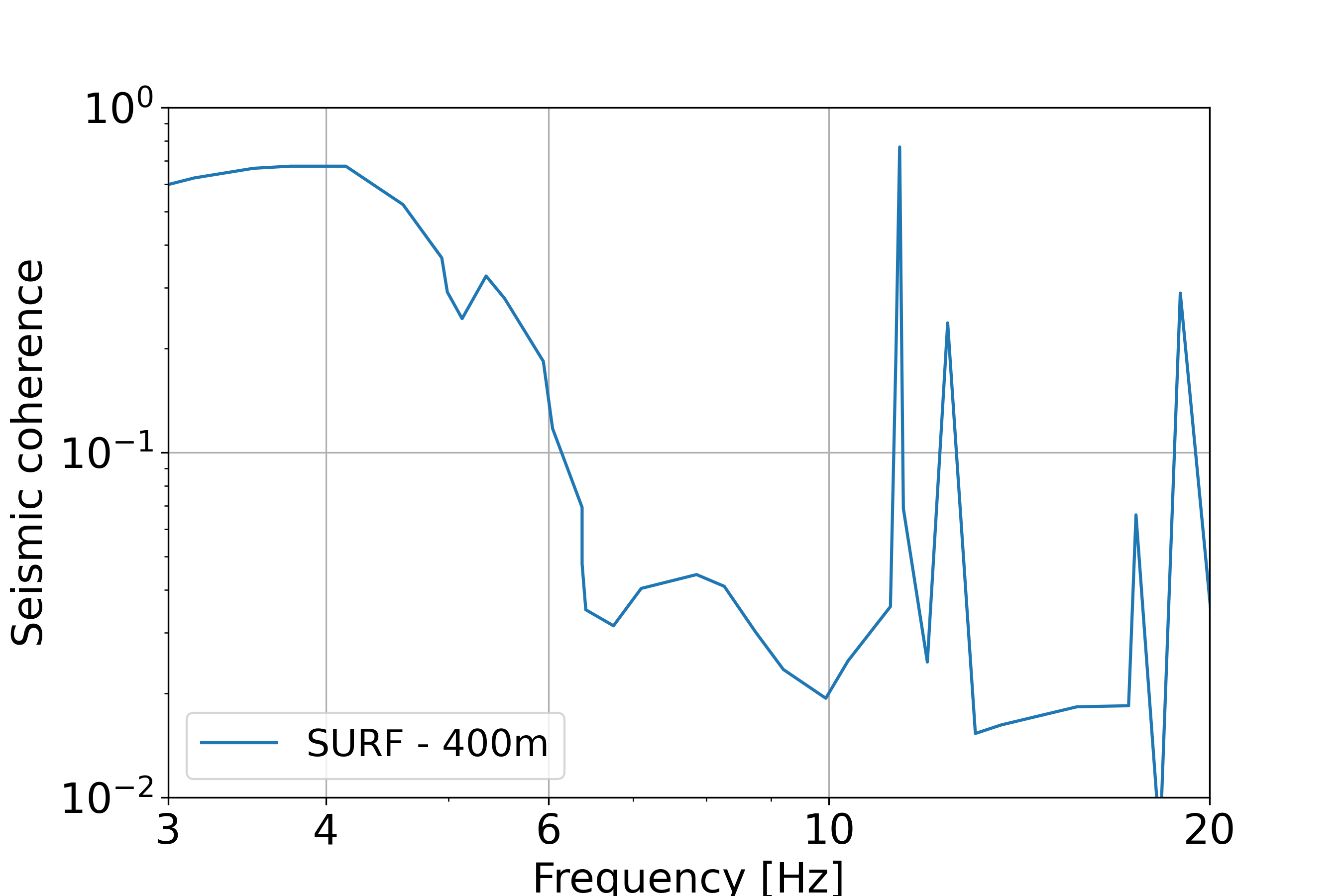}
    \caption{Example measurement of seismic coherence at the Sanford Underground Research Facility \cite{JaEA2024}. The separation between the two seismometers was 400\,m. Horizontal displacement measurements along the same direction were used to calculate the coherence.}
    \label{fig:seismic}
\end{figure}

\begin{table*}[ht!]
    \centering
    \tabcolsep=5pt
    \renewcommand{\arraystretch}{1.5}
    \begin{tabular}{|p{6cm}|p{6cm}|}
    \hline
       {\bf Coherence of environmental forces on test masses in the triangle} & {\bf Coherence between detectors \newline (1-2, 2-3, 3-1)} \\
    \hline
    Scenario 1 & \\
    \includegraphics[height=1.5cm]{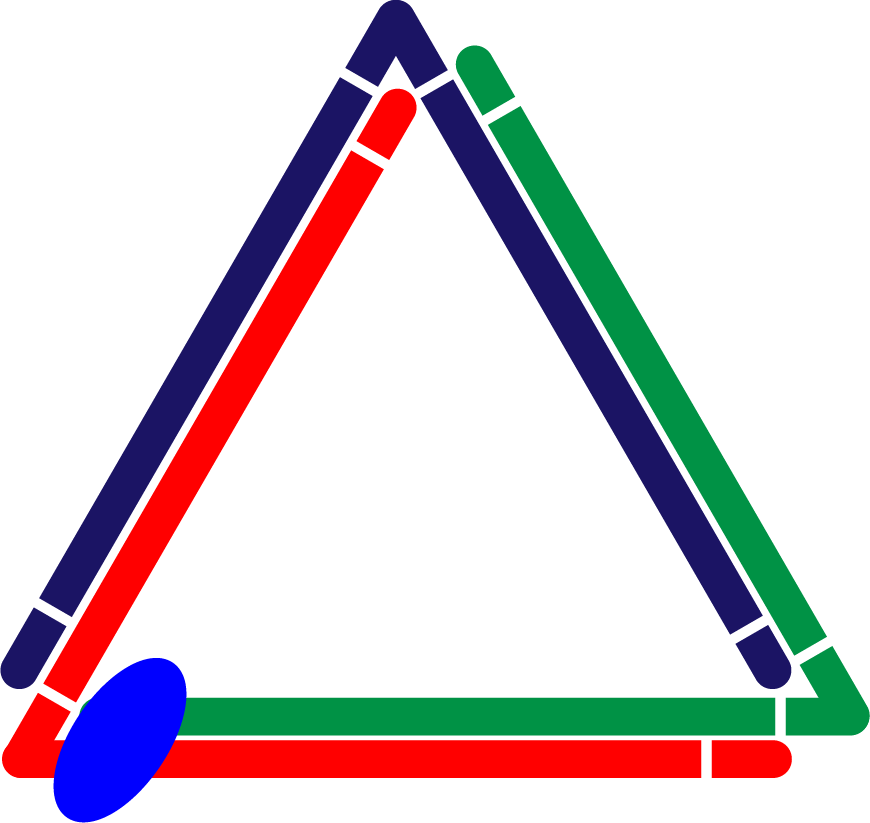} \hspace*{5pt}
    \begin{minipage}[t]{4.5cm}
    \vspace*{-3\baselineskip} \flushleft $\langle F_1^{11}|F_2^{22}\rangle$=1, all others 0
    \end{minipage} & \vspace*{-35pt} $(-0.25,0,0)$ \\
    \hline
    Scenario 2 & \\
    \includegraphics[height=1.5cm]{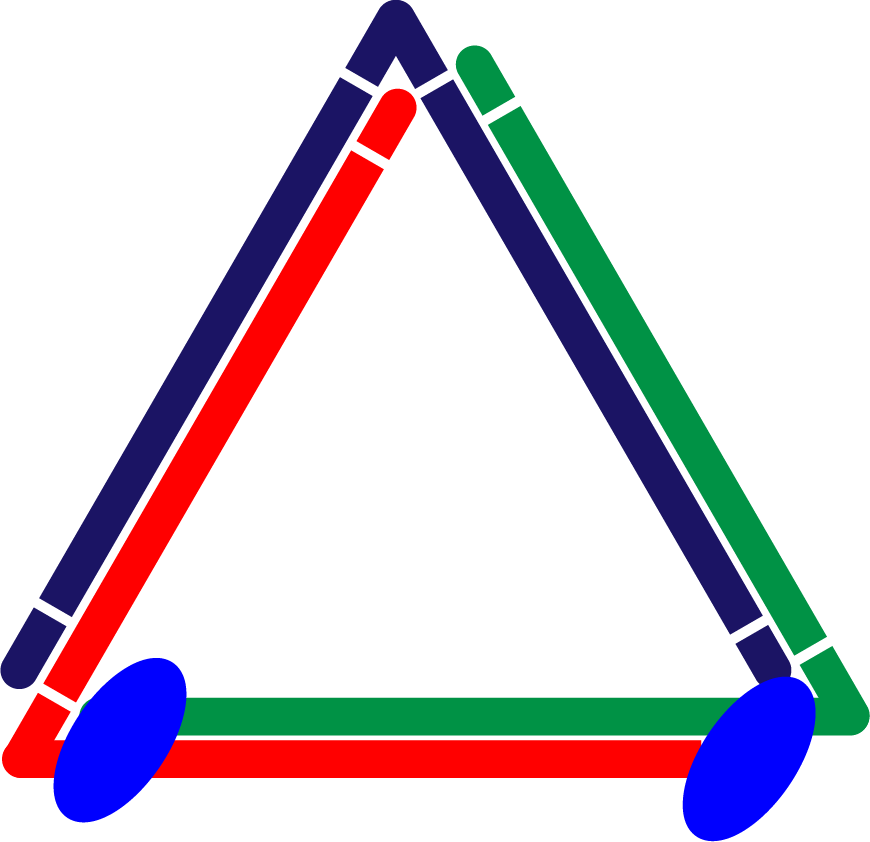} \hspace*{5pt}
    \begin{minipage}[t]{4.5cm}
    \vspace*{-3\baselineskip} \flushleft $\langle F_1^{12}|F_2^{21}\rangle$=$\langle F_1^{11}|F_2^{22}\rangle$=1, all others 0
    \end{minipage} & \vspace*{-35pt} $(-0.5,0,0)$ \\
    \hline
    Scenario 3 & \\
    \includegraphics[height=1.5cm]{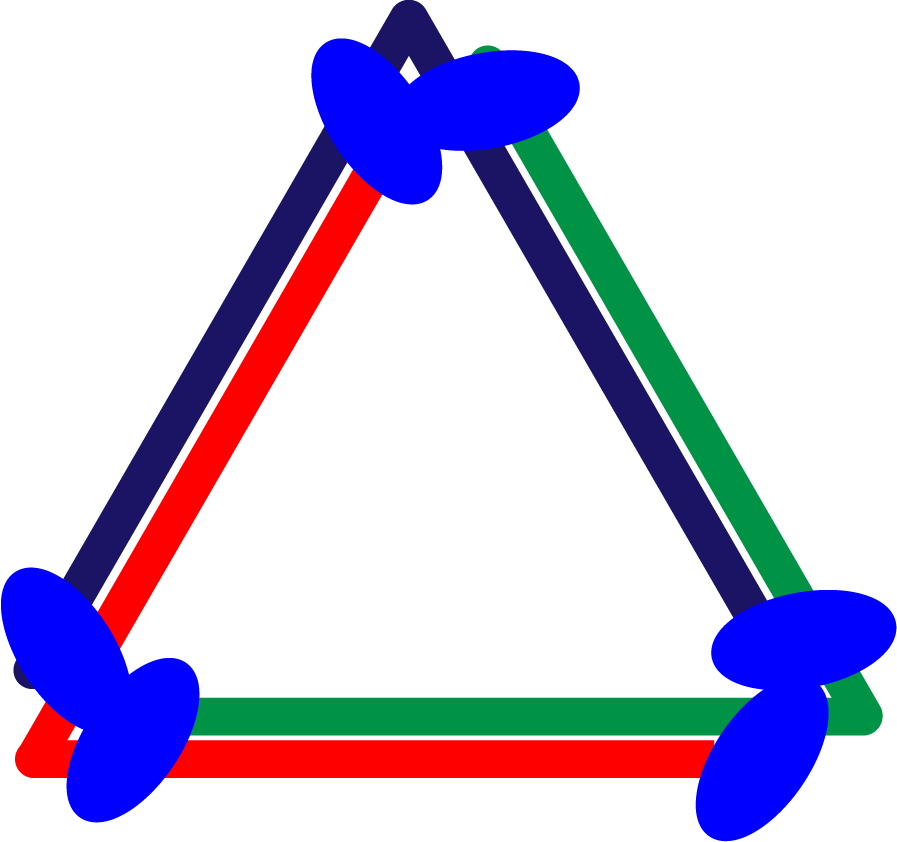} \hspace*{5pt}
    \begin{minipage}[t]{4.5cm}
     \vspace*{-4\baselineskip} \flushleft $\langle F_1^{12}|F_2^{21}\rangle$=$\langle F_1^{11}|F_2^{22}\rangle$=1, $\langle F_2^{12}|F_3^{21}\rangle$=$\langle F_2^{11}|F_3^{22}\rangle$=1, $\langle F_3^{12}|F_1^{21}\rangle$=$\langle F_3^{11}|F_1^{22}\rangle$=1, all others 0  
     \end{minipage} & \vspace*{-35pt} $(-0.5,-0.5,-0.5)$ \\
     \hline
    Scenario 4 & \\
    \includegraphics[height=1.5cm]{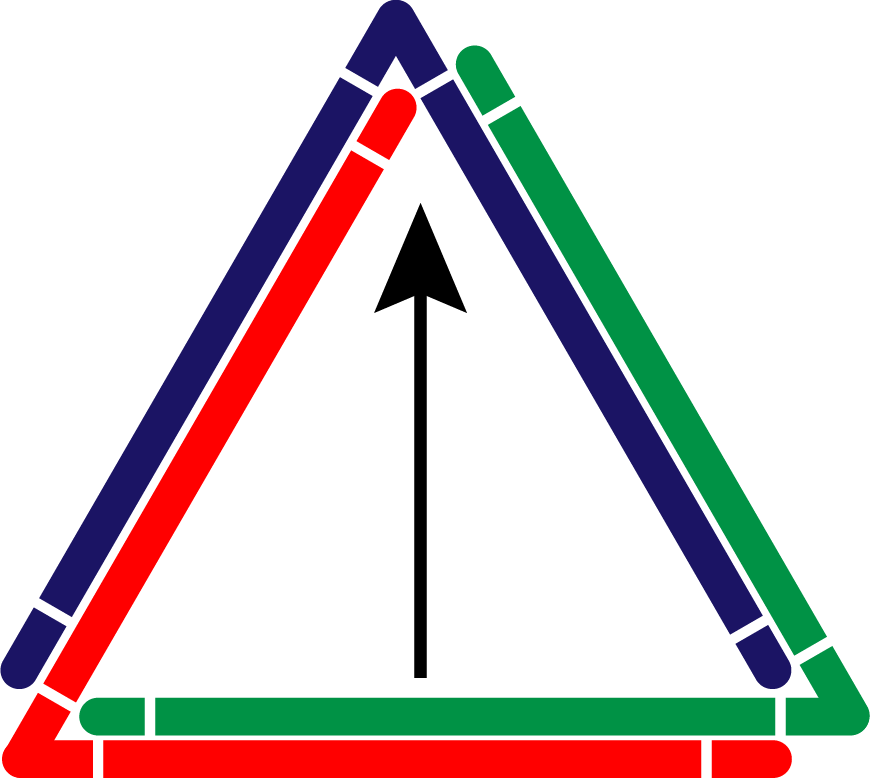} \hspace*{5pt}
    \begin{minipage}[t]{4cm}
     \vspace*{-4\baselineskip} \flushleft common force fluctuations on all test masses  
     \end{minipage} & \vspace*{-35pt} $(0,0,0)$ (no environmental noise) \\
    \hline
    Scenario 5 & \\
    \includegraphics[height=1.5cm]{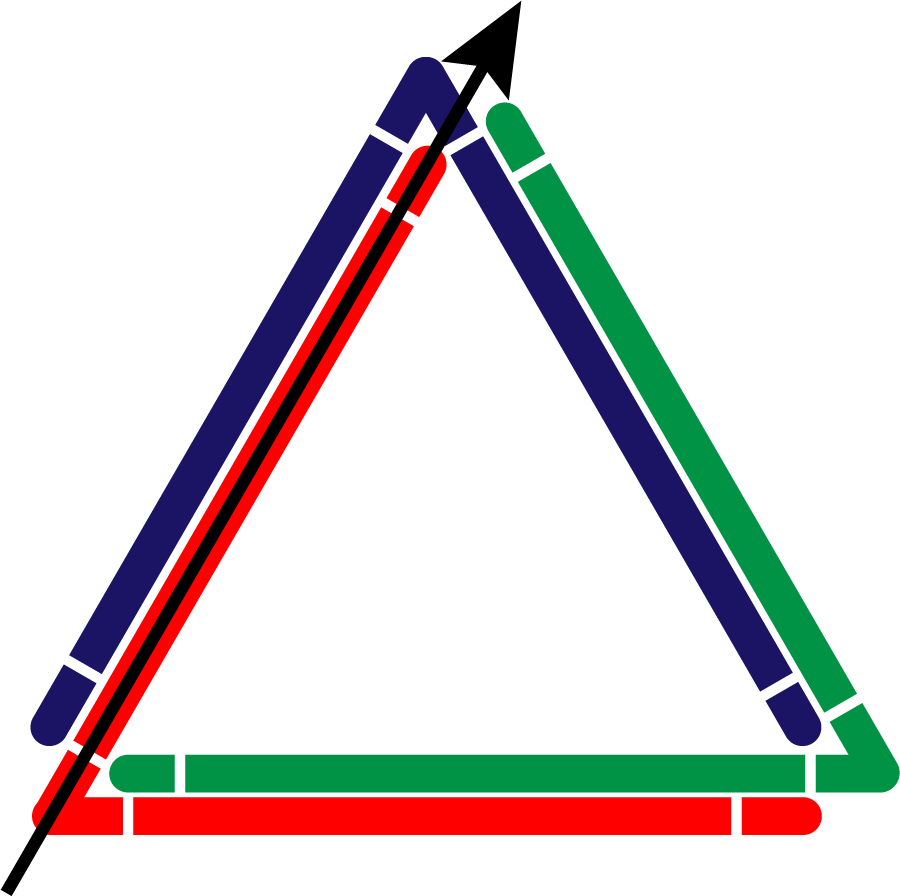} \hspace*{5pt}
    \begin{minipage}[t]{4cm}
     \vspace*{-4\baselineskip} \flushleft common force fluctuations across two vertices
     \end{minipage} & \vspace*{-35pt} $(-1/6,-1/6,1/8)$ \\
    \hline
    \end{tabular}
    \caption{Coherence scenarios of the environmental field (left column) and resulting coherence between detectors (right column) assuming that the correlated noise is identical at all three vertices and dominant over other detector noise. $F_k^{ij}$ denotes the force acting on test mass $k=$detector, $i=$arm, $j=$input or end.}
    \label{tab:correl}
\end{table*}
The first scenario in the table represents perfectly correlated gravitational fluctuations across a single pair of nearby test masses. A physical model of coherence depends on the mix of body-wave polarizations, speeds of seismic waves, the level of anisotropy of the field, and the distance and relative orientation of test masses \cite{DHA2012,Har2019,HaEA2020,BHR2024}. Adopting the simplistic assumption that the compressional waves contribute a fraction $p=0.3$ to the seismic PSD, and using a compressional-wave speed $\alpha=$6\,km/s and shear-wave speed $\beta=3.5$\,km/s, we can calculate the seismic coherence across a distance $D=400\,$m using:
\begin{equation}
\begin{split}
    \gamma_{\rm P} &= j_0(2\pi f D/\alpha)-2j_2(2\pi fD/\alpha), \\
    \gamma_{\rm S} & = j_0(2\pi fD/\beta)+j_2(2\pi fD/\beta), \\
    \gamma_{\rm seis} &=p \gamma_{\rm P}+(1-p)\gamma_{\rm S},
\end{split}
\label{eq:coherence}
\end{equation}
where $\gamma_{\rm P}$ is the coherence of the compressional-wave field, and $\gamma_{\rm S}$ is the coherence of the shear-wave field.
\begin{figure}
    \centering
    \includegraphics[width=0.9\columnwidth]{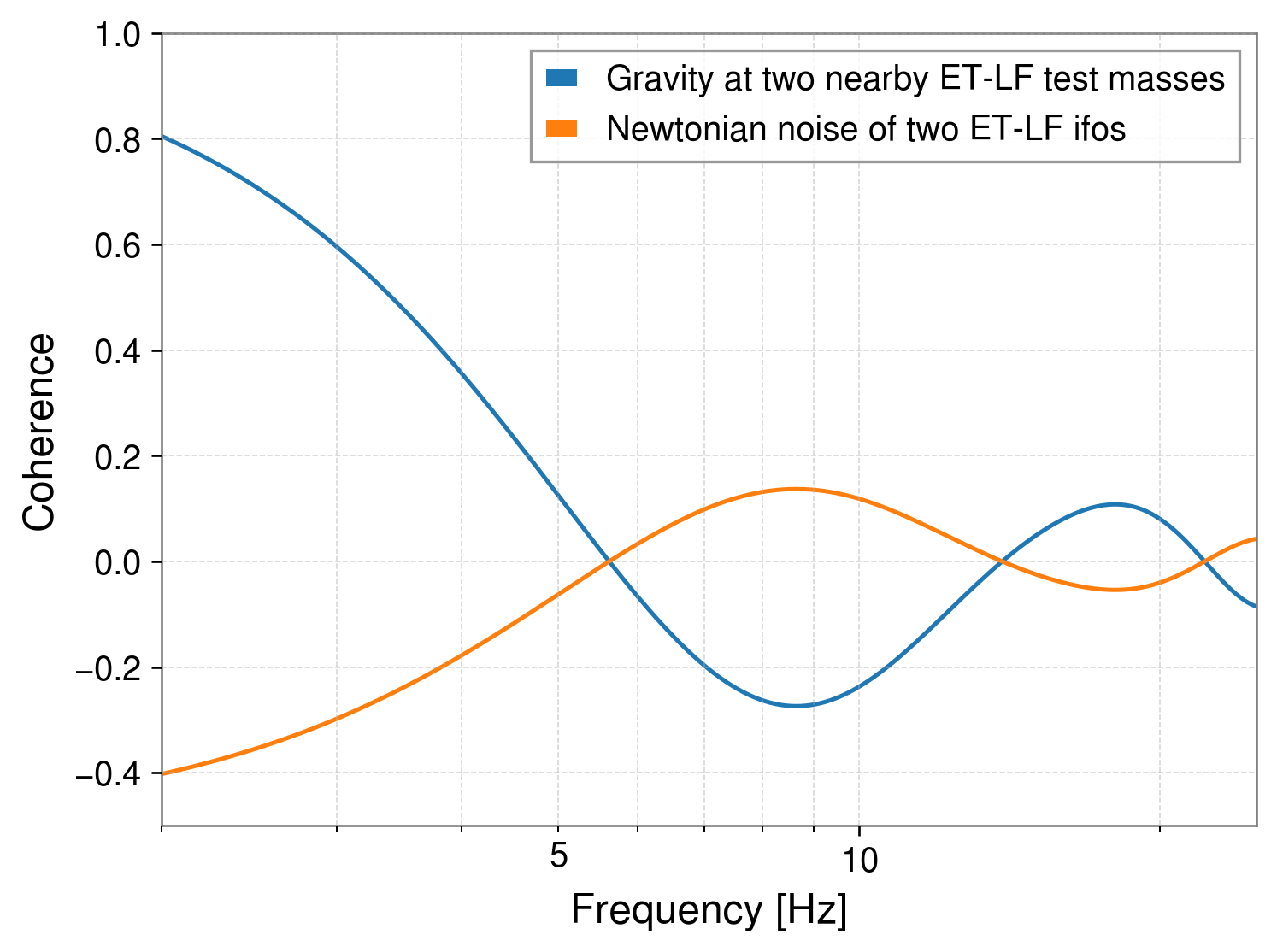}
    \caption{Model of coherence of gravitational fluctuations at the two nearest test masses of two different ET-LF interferometers and coherence of the corresponding Newtonian noise using equations \eqref{eq:coherence} and \eqref{eq:cohifo}. The distance between the two test masses producing the coherence is assumed to be $D=400$\,m, which is approximately what the current optical layout of the ET-LF triangle foresees.}
    \label{fig:nncorrelation}
\end{figure}
Under the assumption of homogeneity of the rock and assuming that the test masses are located at least a seismic wavelength below the surface, the coherence of the gravitational fluctuations and of the associated seismic displacement are calculated from the same seismic coherence functions \cite{Har2019}. This allows us to apply seismic coherence theory to the corresponding gravitational fluctuations; albeit the polarization mix generally has a different impact on seismic and gravitational coherence \cite{BaHa2019}. While from a theoretical perspective it would be correct to use the gravitational coherence model with its distinct mix of shear and compressional-wave coherence, we choose to proceed with the basic seismic coherence of equation \eqref{eq:coherence} with the result shown in figure \ref{fig:nncorrelation} because it matches better the coherence measured at the Sanford Underground Research Facility at a depth of 1.5\,km shown in \ref{fig:seismic}.
Up to 6\,Hz, the measured seismic coherence is similar to what the simple theoretical model predicts. Above 6\,Hz, the observed coherence is weaker than the model predicts, except for a few narrow-band features between 10--20\,Hz likely caused by local machines. Therefore, the gravitational coherence model we adopt in this paper results in a lower coherence above 6\,Hz with respect to the simple theoretical models valid for a homogeneous and isotropic seismic field. 

Figure \ref{fig:nncorrelation} also contains a simplified conversion of gravitational coherence at the two test masses into detector coherence, 
\begin{equation}
    \gamma_{\rm ifo}=-0.5\gamma_{\rm seis}.
    \label{eq:cohifo}
\end{equation}
The minus sign means that the correlation is formed between a test mass in the X-arm of one interferometer and a test mass in the Y-arm of another interferometer. The factor 0.5 means that the correlation only affects one interferometer arm. A more accurate modeling of correlations between ET detectors due to Newtonian noise needs to consider contributions from other test-mass pairs located within a vertex \cite{BHR2024}. 

\begin{figure}
    \centering
    \includegraphics[width=0.9\columnwidth]{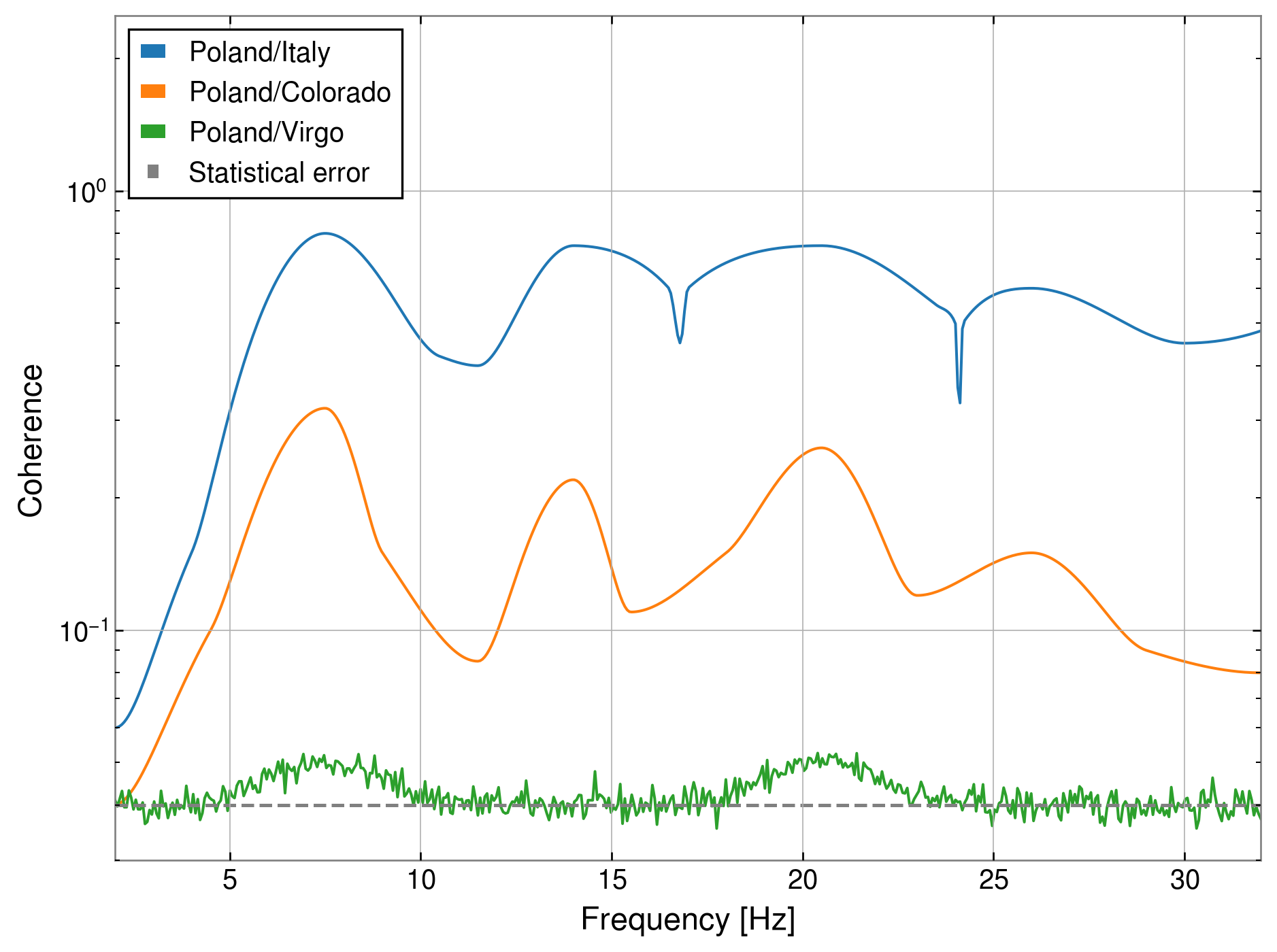}
    \caption{Example measurement of magnetic coherence between distant stations. The orientation of the magnetometers was not optimized to maximize coherence. The results were first reported in \cite{CoEA2018c}. Coherence drops below 5\,Hz in the blue and orange curves because the spectrum of the magnetic fluctuations falls under the magnetometer readout noise. The coherence with the magnetic field at Virgo is low at all frequencies because of local magnetic disturbances. }
    \label{fig:magnetic}
\end{figure}

The last two scenarios in the table require coherent forces across a distance of 10\,km, as could be produced by ambient magnetic fluctuations, see \cite{TCS2013,AtEA2016,KoEA2017,CoEA2018c}, coupling with the magnetic moment of coil-actuator components on the payloads \cite{CiEA2018,CiEA2019}. An example measurement demonstrating the high values of coherence across a distance of about 1200\,km between magnetometers in Italy and Poland is shown in figure \ref{fig:magnetic}.
This means that the coherence of natural magnetic fluctuations measured in a certain direction between two points separated by 10\,km is expected to be very high and represented by scenarios 4 and 5 in table \ref{tab:correl}. The respective detector coherence values are relatively low or even vanishing. This is due to common-mode rejection, which means that noise produced identically at the two test masses of a detector arm cancel out in these two scenarios, because it does not lead to a change in the length of an arm. Also here, a realistic scenario is more complicated since the coupling of a magnetic field at different test masses is expected to be different. Especially the vacuum chambers can locally distort the magnetic field, which has been investigated for the Virgo detector \cite{CiEA2018}. In this paper, we model coherence of magnetic noise between ET-LF interferometers with a frequency-independent value of -0.3.

\section{The impact of correlated noise on GW detection with the triangle configuration}
\label{sec:corrimpact}
In this section, we will apply the formalism developed in \cite{PrEA2002,WoLi2022} to physical models of correlated environmental noise. Noise correlations are described by the off-diagonal elements of the hermitian cross power spectral-density (PSD) matrix $\mathcal S$, e.g., for the $\Delta$-configuration, we have a $3\times 3$ matrix,
\begin{equation}
    \mathcal S=\left(
    \begin{matrix}
        S_{11} & S_{12} & S_{13} \\
        S_{12}^* & S_{22} & S_{23} \\
        S_{13}^* & S_{23}^* & S_{33}
    \end{matrix}\right),
    \label{eq:corrmat}
\end{equation}
where the diagonal elements are the instrument-noise PSDs of the three detectors forming the triangle. The $*$ denotes the complex conjugate. The matrix elements are all frequency dependent.

In order to investigate the issue of noise correlations for the $\Delta$-configuration, we will simplify the noise model. The conclusions do not change if we continued with the generic noise matrix, but it would make the equations more complicated. What we will assume is that all three detectors have the same sensitivity, $S_{11}=S_{22}=S_{33}\equiv S$, and the correlations between all three detectors are the same and real-valued, $S_{12}=S_{13}=S_{23}\equiv \alpha S$. The cross PSD matrix then takes the form
\begin{equation}
    \mathcal S=S\left(
    \begin{matrix}
        1 & \alpha & \alpha \\
        \alpha & 1 & \alpha \\
        \alpha & \alpha & 1
    \end{matrix}\right).
    \label{eq:corrmatex}
\end{equation}
A possible approach to describe the impact of the correlations is to switch into the noise eigenbasis, i.e., we form linear combinations of the GW channels of the three detectors so that their noise is uncorrelated. The orthonormal eigenvectors of the matrix in equation (\ref{eq:corrmatex}) are given by
\begin{equation}
    \begin{split}
        \vec c_1 &= (-1,0,1)/\sqrt{2},\\
        \vec c_2 &= (-1,2,-1)/\sqrt{6},\\
        \vec c_3 &= (1,1,1)/\sqrt{3}.
    \end{split}
    \label{eq:eigenvec}
\end{equation}
The corresponding eigenvalues are
\begin{equation}
    \begin{split}
        \lambda_1 &= 1-\alpha,\\
        \lambda_2 &= 1-\alpha,\\
        \lambda_3 &= 1+2\alpha.
    \end{split}
\label{eq:eigenval}
\end{equation}
While in general, the environment will cause complex-valued correlations, we have shown in section \ref{sec:correl} that the dominant contributions have a negative sign in the case of Newtonian and magnetic noise. Consequently, the noise in the eigenbasis is enhanced in two channels compared to the noise PSD of an interferometer, and  decreased in the third channel. Equation \eqref{eq:eigenval} shows that $\alpha$ must be larger than -0.5 or the third eigenvalue would become negative, which is impossible for a noise PSD. The model presented in section \ref{sec:correl} is consistent with this result. 

To understand the impact of noise correlations on GW detection, we must invoke the null stream. As can be seen from equation \eqref{eq:eigenvec}, the third eigenvector has three identical components, which means that it describes a linear combination of detector channels corresponding to the sum in equation (\ref{eq:nullstream}), i.e., $\vec c_3$ forms the null stream. We just argued that for $\alpha<0$, it is exactly this channel where noise is reduced while the other two channels of the eigenbasis have increased noise. The potential benefit from noise correlations goes away since the noise reduction is achieved in the same channel where the GW signal is not present. 

We can calculate the total SNR of the three detectors in the eigenbasis normalized by the total SNR in the absence of noise correlations ($\alpha=0$):
\begin{equation}
    \frac{\sum\limits_{k=1}^3{\rm SNR}^2_k(\alpha)\bigg|_{\rm eigen}}{\sum\limits_{k=1}^3{\rm SNR}^2_k(\alpha=0)}=\frac{1}{1-\alpha}.
\label{eq:snrcorr}
\end{equation}
For example, if the correlations assume the minimal value $\alpha=-0.5$ of the model in equation (\ref{eq:corrmatex}), then the SNR of the $\Delta$-configuration would be reduced by a factor $\sqrt{2/3}\approx 0.82$. 

\begin{figure}[ht!]
    \centering
    \includegraphics[width=0.95\columnwidth]{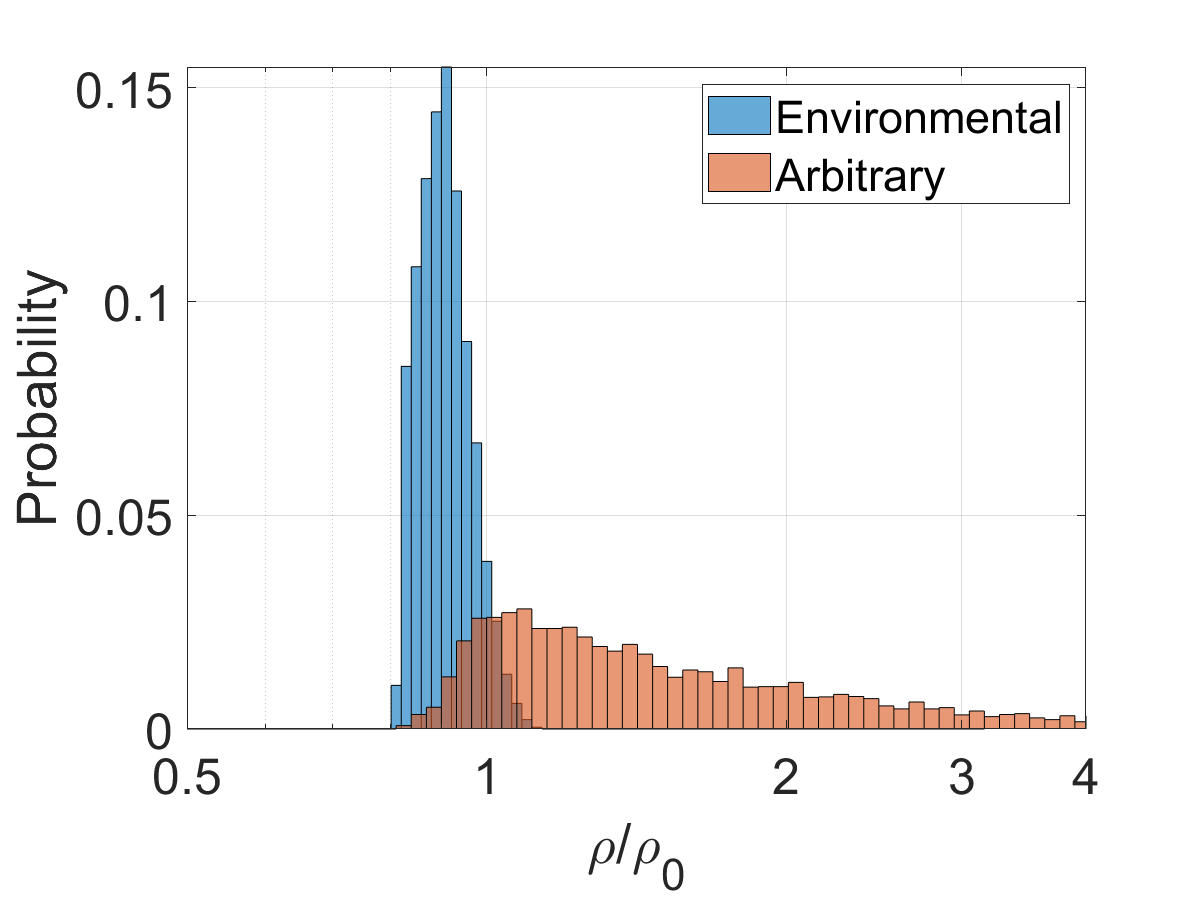}
    \caption{Distribution of SNRs when randomly drawing three (generally complex-valued) coherence values between the three detectors of the $\Delta$-configuration. The SNRs are evaluated below 6\,Hz where Newtonian noise is strongly correlated. The effect on SNRs above 6\,Hz is negligible since seismic coherence is low.}
    \label{fig:corrsnr}
\end{figure}

One might wonder if the simplified form of the cross-PSD shown in eq.~\eqref{eq:corrmatex} biases our conclusions. To investigate this point, we now allow for arbitrary, but physically possible, correlations between the three detectors. Figure \ref{fig:corrsnr} shows the values of the normalized SNR analogous to eq.~\eqref{eq:snrcorr} drawing coherence values with arbitrary complex phases and absolute values between 0 and 1 keeping those combinations that lead to a positive definite cross-PSD matrix. The distribution of SNRs is compared with the case when coherence values are consistent with an environmental origin, i.e., assuming that $\alpha\in [-0.5,\,1]$ and complex phases limited to $\phi\in[0,2\pi Df_{\rm max}/\beta]$, where $\beta=3.5\,$km/s is the speed of seismic shear waves (the slowest waves considered in this study), $D=400\,$m is the distance between the closest test masses of two ET-LF interferometers, and $f_{\rm max}=6$\,Hz.
We find that the distribution of normalized SNRs with environmental correlations mostly lie below 1, i.e., environmental correlations are detrimental to GW detection, while in the arbitrary case, SNRs can be high. Such a benefit of correlated noise represents a hypothetical case where, for example, test-mass actuators are used to inject noise with "unnatural" phase relations between different detectors.

\begin{figure}[ht!]
    \centering
    \includegraphics[width=0.95\columnwidth]{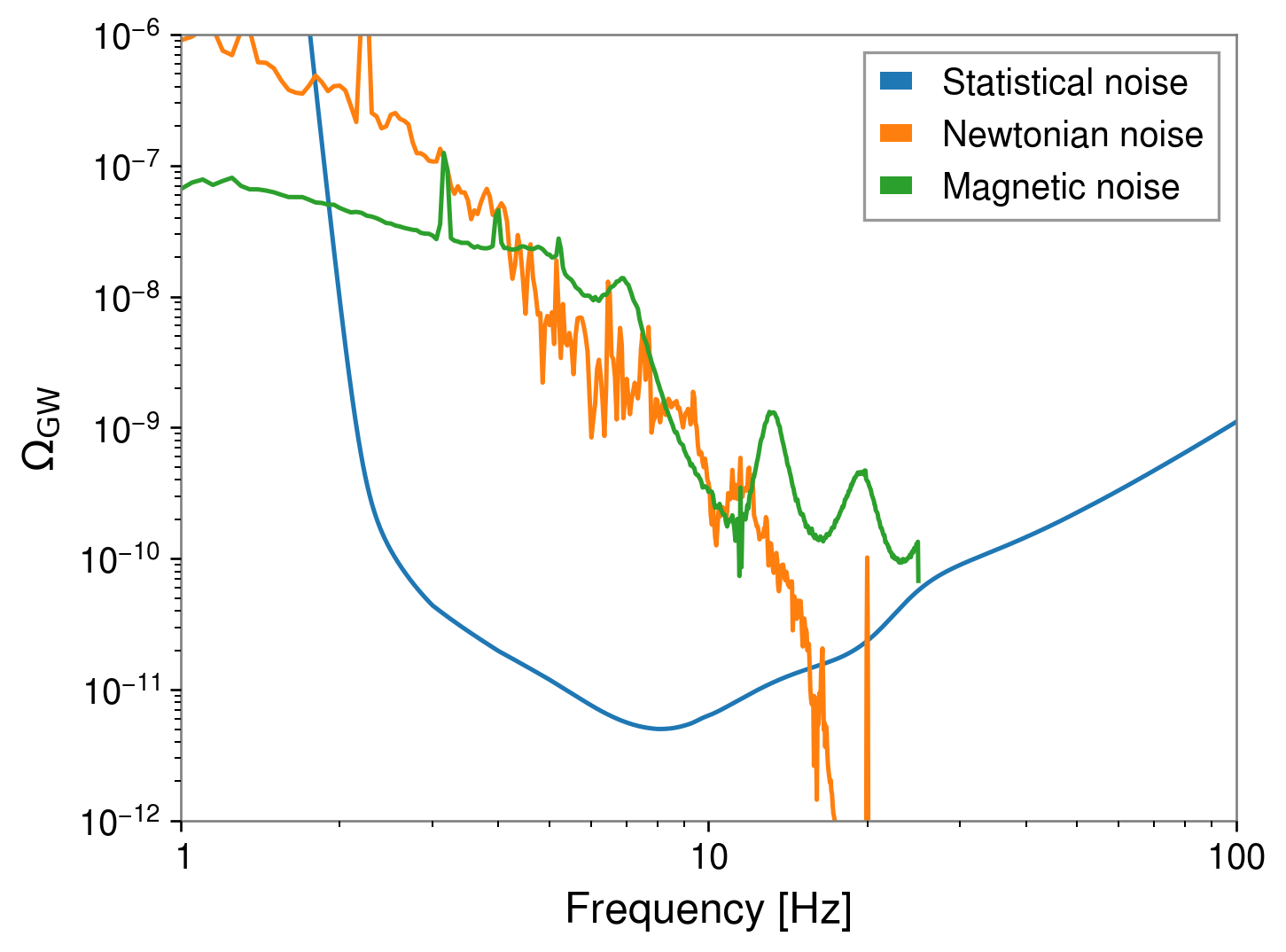}
    \caption{Correlated environmental noise poses a sensitivity limitation to stochastic GW searches. The correlated noise is shown without mitigation.}
    \label{fig:omega}
\end{figure}

Correlated noise also limits the sensitivity of ET to stochastic GW backgrounds \cite{Caporali2025}. Without noise correlations, stochastic searches are limited by the statistical errors of correlation estimates. In units of relative energy density, the statistical error is given by \cite{ShHa2020}
\begin{equation}
\Omega_{\rm stat} = \frac{10\pi^2f^3}{3H_0^2\gamma_{12}(f)}\sqrt{\frac{S_n^1(f)S_n^2(f)}{N}},
\end{equation}
where $H_0$ is the Hubble constant, $\gamma_{12}(f)$ is the overlap reduction function between two detectors of the $\Delta$-configuration, $S_1(f),\,S_2(f)$ are the PSDs of uncorrelated noise of the two detectors, and $N$ is the number of averages that go into the correlation estimate. Instead, the limitation from correlated noise expressed in terms of the cross-PSD $S_n^{12}(f)$ is given by
\begin{equation}
\Omega_{\rm corr} = \frac{10\pi^2f^3}{3H_0^2\gamma_{12}(f)}|S_n^{12}(f)|.
\end{equation}
Figure \ref{fig:omega} shows the statistical error assuming one year of observation, and the limitation from correlated environmental noise calculated from simulated time series of two of the detectors. Here, the Newtonian noise is modeled using data from the Terziet borehole in the Netherlands from December 16, 2019 \footnote{https://ds.iris.edu/mda/NL/TERZ/?starttime=2019-06-03\&endtime=2599-12-31}, while the magnetic-noise model is based on magnetometer data from the Sos Enattos underground station using the coupling function observed in Virgo \cite{CiEA2018}, which is similar to the coupling observed in LIGO at frequencies up to 20\,Hz despite the different payload designs \cite{Janssens2025}. The coupling function is rescaled to the larger test masses in ET. Note that we take the absolute value of correlations for the stochastic GW detection. When searching for an isotropic GW background, GW correlations between the detectors are real-valued, which means that the real part of the correlations can be used potentially leading to a phase-dependent sensitivity gain. Here we cannot use it because we do not have enough measurement stations at the candidate sites to accurately model the phase of the environmental-noise cross-PSD (see next section for details of the time-series simulation), which strongly depends on how noise from different vertices combines. Instead, the absolute value is modeled with relatively minor inaccuracies.

In summary, correlated noise has an important impact on the observational capabilities of the $\Delta$-configuration especially with respect to stochastic searches. The detrimental effect of correlated noise can be mitigated with technologies like magnetic shielding and Newtonian-noise cancellation. Noise mitigation will reduce correlated noise, but with respect to magnetic noise and Newtonian noise, the functionality of the mitigation systems has not been tried and proven in existing GW detectors so far, which leaves question marks and risks concerning their effectiveness. The null stream is of no use here since it results from a change of basis without providing additional information to disentangle the contributions to the cross-PSD matrix of the $\Delta$-configuration \cite{WoLi2022,Janssens2023}. 

\section{Transient noise}
\label{sec:transient}
In the previous sections, we modeled noise correlations in the ET triangle, which manifest themselves as average sensitivity limitations reducing the SNR of transient as well as of long-lasting GW signals. A similar situation is to be expected with respect to detector-noise transients also known as glitches. A recent study of glitches in the LIGO detectors highlights the role of the environment glitches \cite{Ferreira2025}. The low-frequency sensitivity improvement of LIGO Livingston between the third and forth observing run caused environmental glitches to be the most numerous glitch family, with more than 10 times as many environmental glitches as LIGO Hanford. Given ET's ambitious low-frequency sensitivity goal and that many detectable GW signals will be present simultaneously at any time, glitches are expected to be a major issue also for ET data analysis \cite{Allocca2021}. Here are a few key points to keep in mind:
\begin{itemize}
    \item ET's low-frequency sensitivity target is up to 6 orders of magnitude better than of current detectors, which increases the glitch rate \cite{Ferreira2025};
    \item Underground construction is expected to mitigate some of the glitches observed in current detectors;
    \item New technologies for ET like improved active seismic isolation \cite{MoMa2019}, inter-platform sensing and control \cite{Koehlenbeck2023,Andric2026}, and nonlinear control techniques \cite{Buchli2025} are under development (also) to reduce glitches.
    \item Every glitch needs to be mitigated or it will affect the analysis of hundreds to a few thousand detectable GW signals present at any time.
\end{itemize}
It is impossible to predict the full glitch population of ET, but there are two families of glitches that can be modeled today based on environmental data from the candidate sites: glitches produced by magnetic and gravitational coupling with the ET-LF payloads. It is possible to model these glitches because the detector has a simple linear response to the corresponding transients in the seismic and magnetic fields \cite{AmEA2020}. 

The main challenge of providing realistic glitch models for the $\Delta$-configuration comes from the limited site data available today. Gravitational and magnetic coupling will produce coincident glitches in the ET-LF interferometers, but not necessarily glitches that have exactly the same shape. The procedure followed here is
\begin{enumerate}
    \item Use environmental data from three collocated orthogonal measurements, i.e., the three channels of a borehole seismometer and orthogonal magnetic measurements. These channels have similar spectra and contain coincident (but not identical) transients, which is precisely what we need to model ET-LF noise.
    \item Transform the environmental data into environmental detector noise. The result is an ET-LF noise model without the correct correlations between interferometers.
    \item Pass the ET-LF environmental noise from the previous step through a filter that implements a frequency-dependent model of correlations between ET-LF interferometers. In practice, this is done by calculating the Fourier transform of the input time series, calculating the Cholesky decomposition $S(f) = L(f)L^\dagger(f)$ of the normalized cross-PSD matrix $S(f)$ of the correlated interferometer noise, multiplying the matrix $L(f)$ to the vector of Fourier-transformed input time series, and calculating the inverse Fourier transform, which yields the correlated time series.
\end{enumerate}
The end product consists of three time series with coincident glitches and environmental noise that is correlated according to a frequency domain model like the one shown in figure \ref{fig:nncorrelation}. It should be noted that this model can only consider the glitches produced at one of the three detector vertices (unless there were environmental data from all three vertices, which there are not today). 

\begin{figure*}[ht!]
    \centering
    \includegraphics[width=0.49\linewidth]{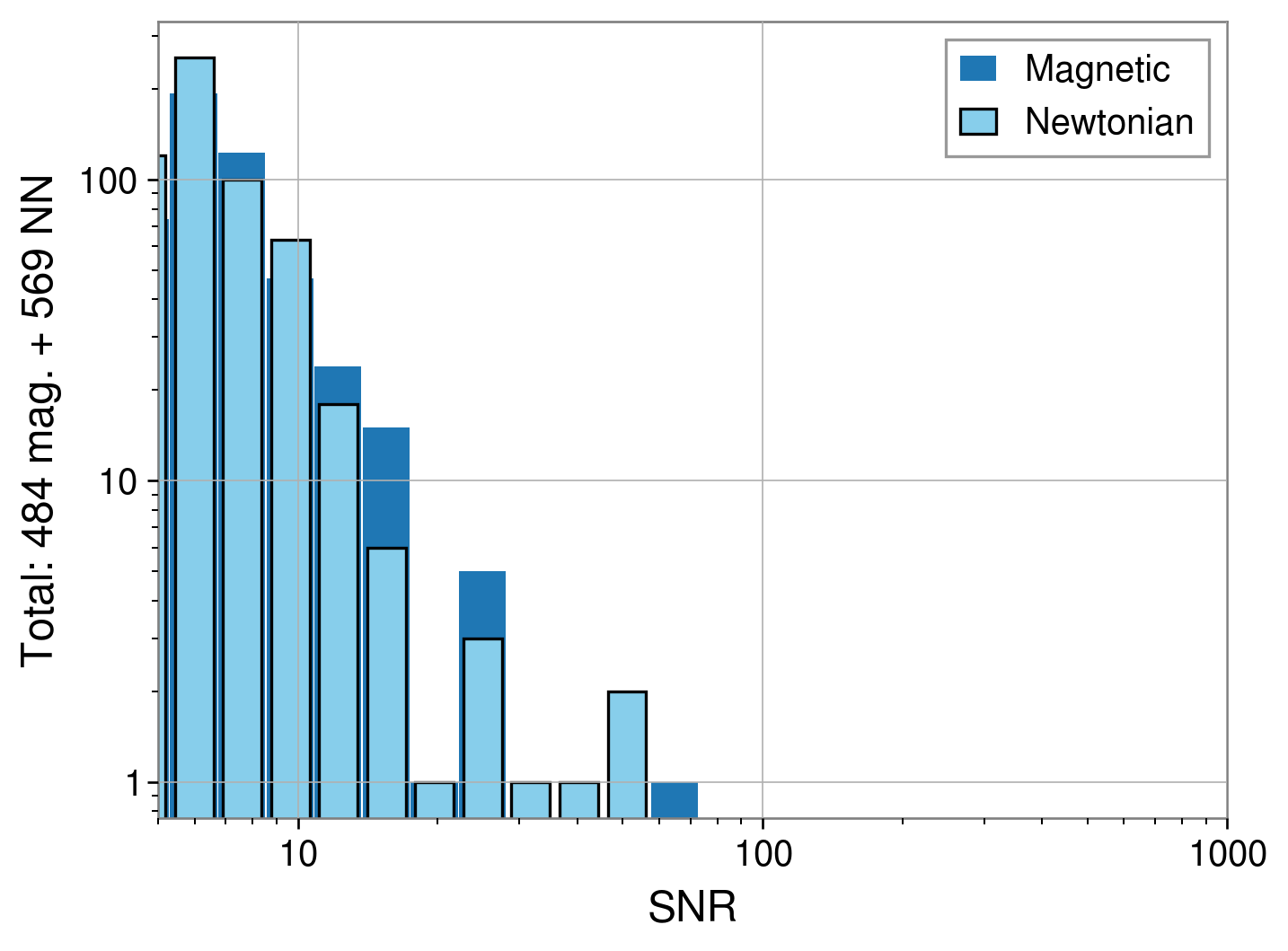}
    \includegraphics[width=0.49\linewidth]{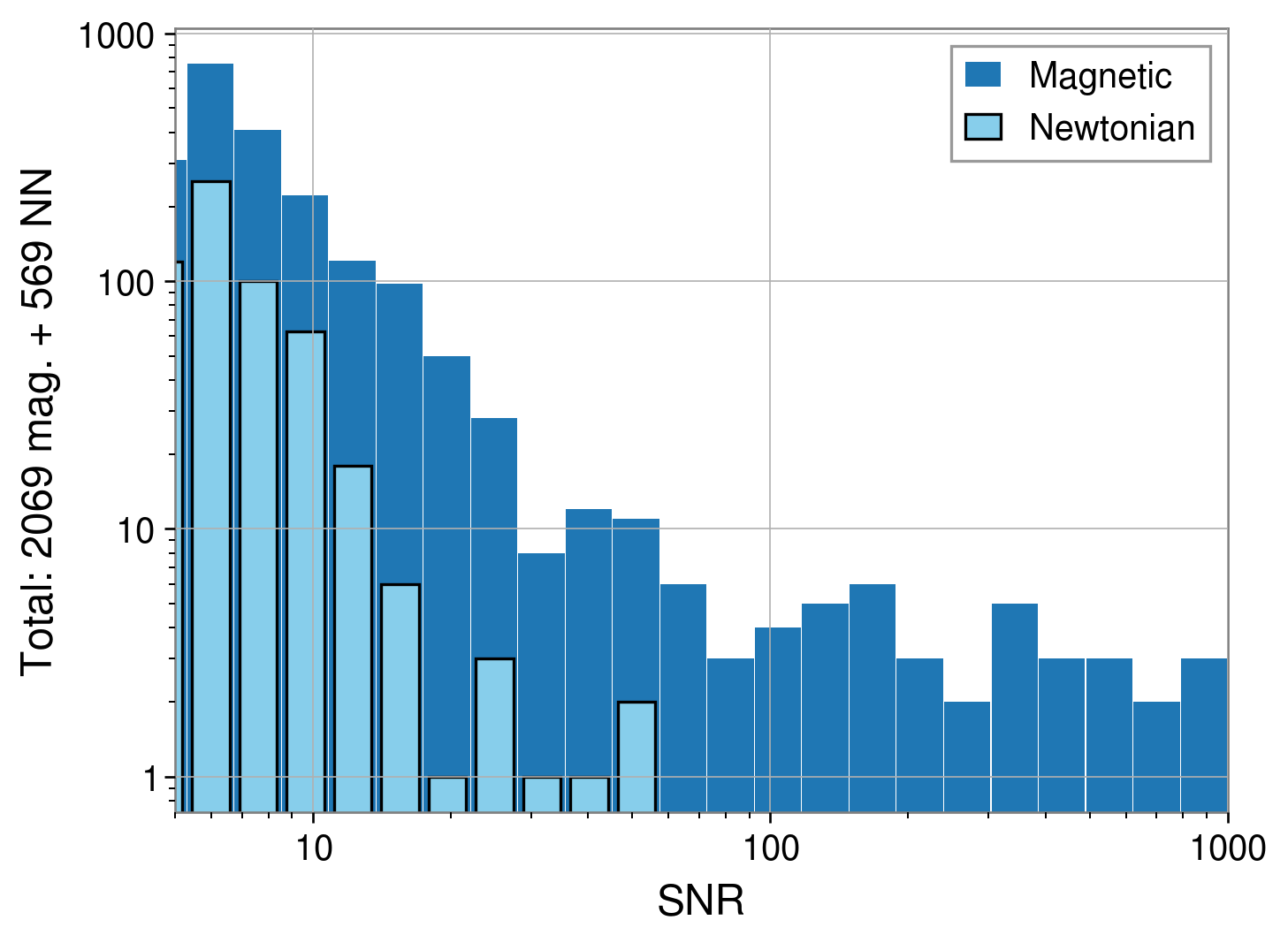}
    \includegraphics[width=0.49\linewidth]{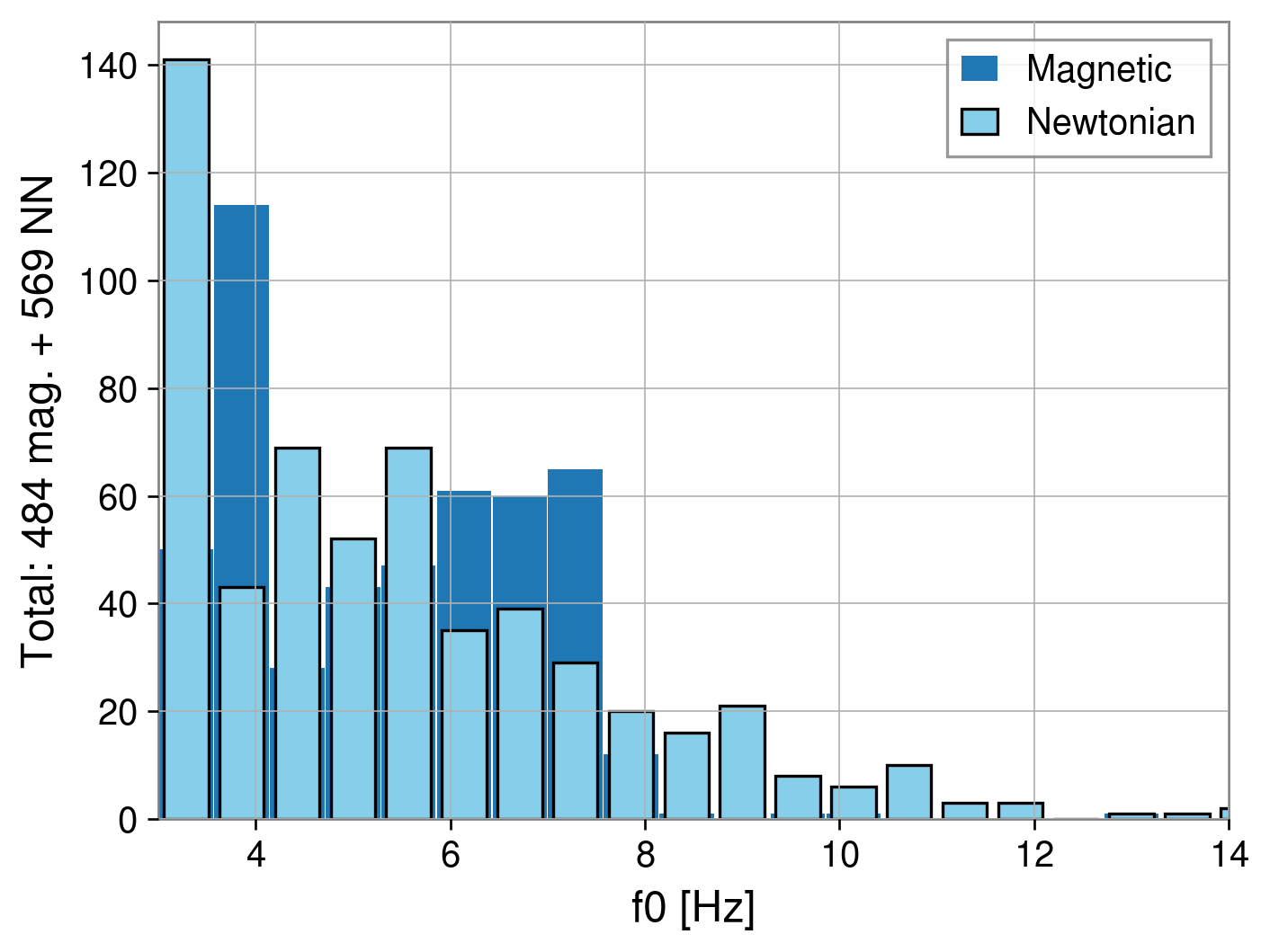}
    \includegraphics[width=0.49\linewidth]{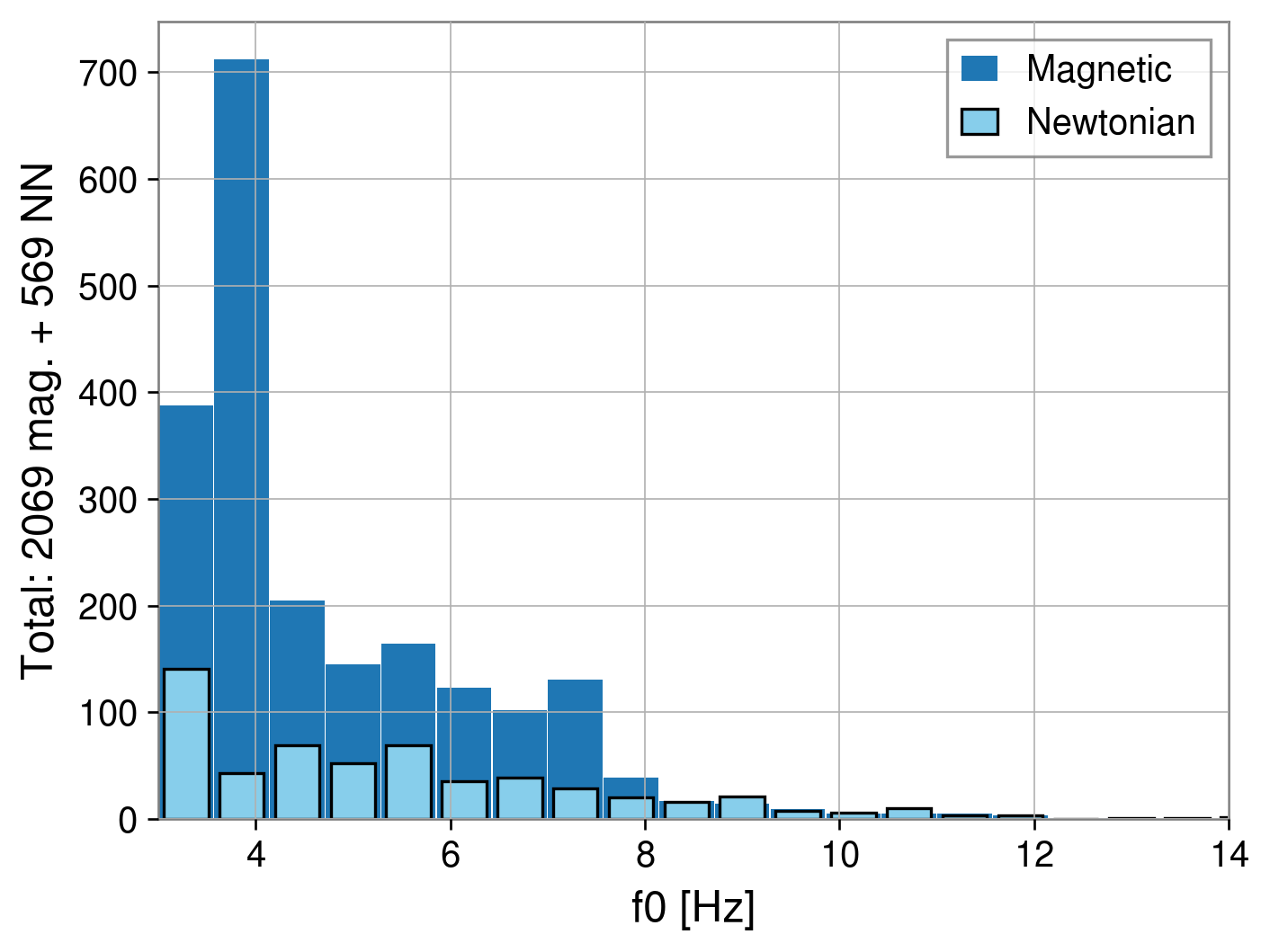}
    \caption{Magnetic and Newtonian glitches detected with the LVK Omicron pipeline using one day of simulated ET data. The upper plots show the distribution of Omicron SNRs, which means the SNRs of glitches with respect to the average noise. The lower plots show the distribution of the characteristic glitch frequencies. The left plots are based on magnetometer data from a typical day, while the right plots are based on data from one day after a thunderstorm passed Sos Enattos. The SNR of the glitches is with respect to the elevated low-frequency PSD due to Newtonian and magnetic noise (see figure \ref{fig:nullspec}).}
    \label{fig:glitchhisto}
\end{figure*}

We use the Terziet borehole seismic data to model Newtonian noise. Instead, high-quality magnetometer data are currently available only from the site in Sardinia, and we use data from an underground measurement station at the former Sos Enattos mine. The ET-LF time series are calculated separately with Newtonian noise and magnetic noise so that the glitches can be analyzed for each case. We use the Omicron trigger search, a standard tool in the LVK collaboration, to find glitches in the simulated data. The results are shown in figure \ref{fig:glitchhisto} for one day of simulated data applying a trigger-SNR threshold of 5. 
The total number of triggers shown in the y-axis labels contains Newtonian as well as magnetic glitches. The left plots represent a typical day, while the right plots are based on data from one day after a thunderstorm passed the site. As mentioned earlier, these triggers originate from the environmental fields at one of the vertices. For the Newtonian glitches, it means that the total number of glitches will likely be three times as high in each ET-LF interferometer. For the magnetic glitches, many will be from distant sources or strong enough to cause glitches at all three vertices, in which case the total number of magnetic glitches is less than three times the number found here. Conservatively assuming that every magnetic glitch is observed at all three vertices, we obtain an estimate of $200,000$ magnetic glitches per year in each ET-LF interferometer. The number of Newtonian glitches per year and interferometer is estimated to be $600,000$.

Glitches can lead to false GW detections \cite{Babak2013,DalCanton2021,Davis2021}, which is more likely to be triggered when glitches appear coincidentally in several interferometers and depending on how many interferometers are online \cite{Allen2005,PhysRevD.102.022004,Cabero_2019,Abbott_2020}, and they are more likely to affect generic burst searches \cite{DrEA2021,Macquet2021} than searches with matched filters \cite{PhysRevD.108.043004,DalCanton2021}. Environmental sensors can be used to veto such false detections. False alarms can also be triggered by Gaussian noise \cite{MoEA2023}, and the fraction of Gaussian false alarms is higher when the GW signal is longer, i.e., glitches are less likely to produce false alarms for long GW signals \cite{Nitz_2017}. The Gaussian trigger distribution as a function of trigger SNR $\rho$ is given by
\begin{equation}
    N(\rho)=N_0\exp(-\rho^2+\rho_0^2),
\end{equation}
where $\rho_0$ is the SNR threshold for a detection. Detections with SNR as low as 8 together with low false-alarm rates are possible today \cite{AbEA2021a}, where the background near threshold is always effectively Gaussian (at least for the modeled searches). For near-threshold triggers, the distribution can be approximated as
\begin{equation}
    N(\Delta\rho)\approx N_0\exp(-2\Delta\rho\rho_0),
\end{equation}
with $\Delta\rho=\rho-\rho_0$, which matches the observation that the SNR distribution of false alarms near threshold can be approximated by an exponential \cite{Nitz_2017,CabournDavies2022}. The rate $N_0$ at threshold depends on the PSD of the detector as well as on the signal waveform \cite{MoEA2023}. This Gaussian distribution of trigger SNRs fits best for the longest GW signals produced by low-mass compact binaries, and it will be a good approximation for all neutron-star binary signals in ET. False alarms will be more commonly produced by (non-Gaussian) glitches when the GW signals are short. 

More important though is the effect on parameter estimation especially on short signals \cite{DaEA2022,PaEA2022,Narola2025}. With hundreds to a few thousand of detectable signals present at any time in ET data, every glitch will need mitigation. Glitch mitigation has fundamental limits. A standard technique is to use parameterized glitch models. BayesWave uses a sum of wavelets to model and subtract a glitch from GW data \cite{AbEA2017d,Cornish2015}. If the glitch model is generic enough, all the glitch energy can be subtracted, but together with it, part of the Gaussian noise is subtracted as well \cite{CuHa2006,HaEA2008,ShHa2020}. This leaves artifacts in the data, which can affect parameter estimation much like an unmitigated glitch. The average SNR of the subtracted Gaussian noise is equal to $\sqrt{N_p}$, where $N_p$ is the number of parameters of the glitch model. Hence, increasing the dimension of the glitch model enhances the artifact, and there is an optimum value for $N_p$.

When a glitch overlaps with a GW signal, glitch and GW-signal modeling can interfere, and the problem is exacerbated by the simultaneous presence of many signals; albeit the forest of GW signals is penetrable with standard analysis techniques in the case of ET \cite{Wu2023}. This is where the null stream was proposed to help with the glitch mitigation, because GW signals are absent in the null stream \cite{Narola2025}. As we will show in section \ref{sec:null}, the null stream does not help with the mitigation of environmental glitches, which constitute the majority of glitches in current detectors, and our predictions of Newtonian and magnetic glitches suggest that this will also be the case in ET despite its underground location. 

\section{Utilizing the null stream to mitigate glitches}
\label{sec:null}
As we have seen in section \ref{sec:nullspace}, the existence of the $\Delta$ null stream is connected to a redundancy of the signal space, which basically comes from the fact that a GW has only two independent polarizations and only two interferometers are needed to measure them. The unique property of the null stream of the $\Delta$-configuration is that it is formed by a simple sum of interferometer data independent of the GW amplitudes. All GW signals vanish in the same null stream; or more accurately, the GW signals are strongly suppressed at low-enough frequencies \cite{Virtuoso2025}.

People considered the null stream to be of potential use for the assessment and mitigation of noise \cite{ReEA2012,GNH2022}. The question to answer is whether the null stream facilitates the modeling of detector noise, be it in the form of cross-PSDs as would be required for stochastic searches, or in the form of glitches, which affect the analysis of generic bursts and compact-binary signals.

\begin{figure}[ht!]
    \centering
    \includegraphics[width=\columnwidth]{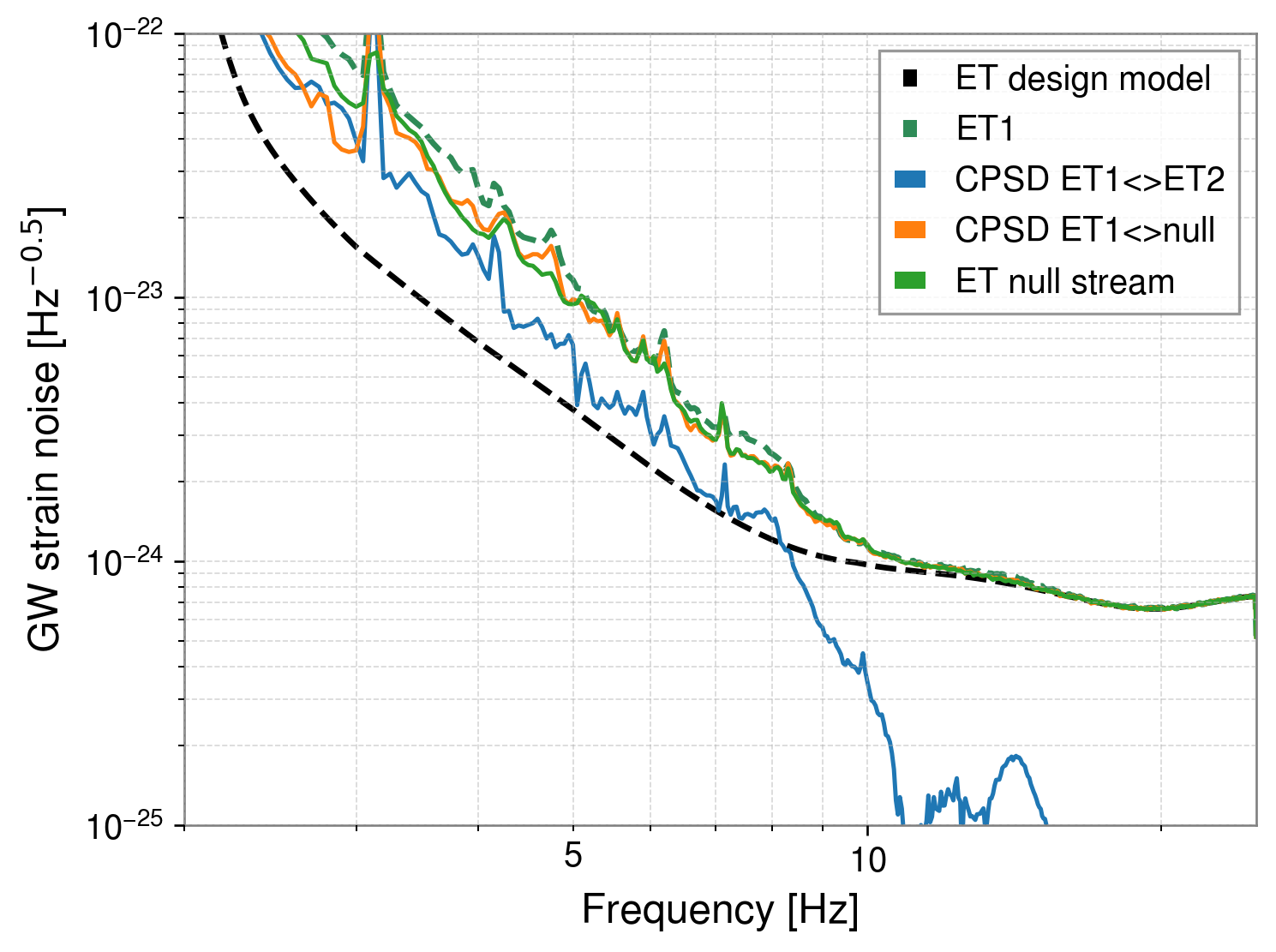}
    \includegraphics[width=\columnwidth]{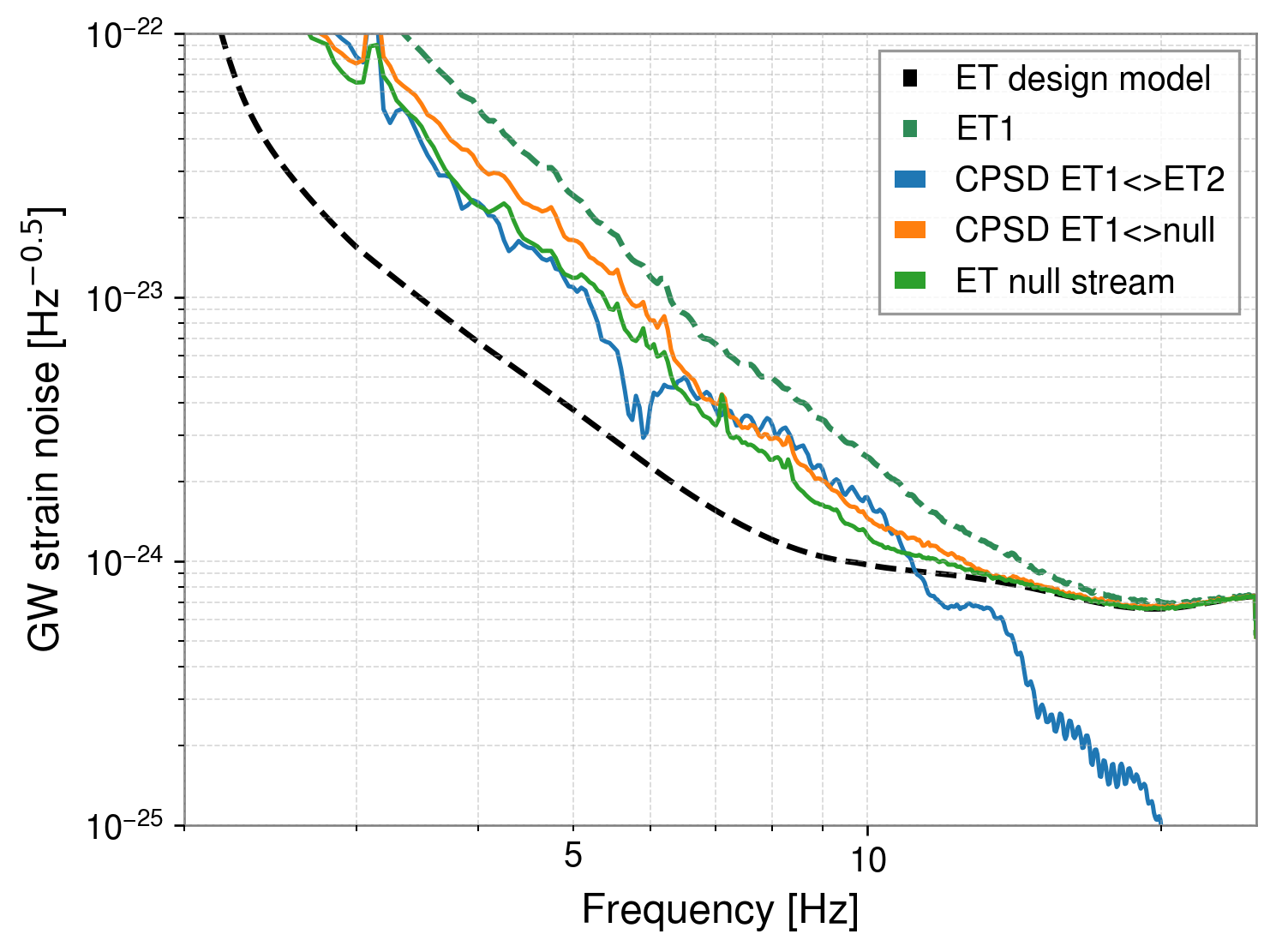}
    \caption{Square roots of the PSDs of ET1, of the null stream, and of the cross-PSDs between ET1 and ET2 and ET1 with the null stream. The upper plot represents a magnetically quiet day, while the lower plot is using magnetometer data one day after a thunderstorm passed the site. The time series are calculated without assuming any magnetic-noise and Newtonian-noise mitigation.}
    \label{fig:nullspec}
\end{figure}

We start with an analysis of the noise spectra. Figure \ref{fig:nullspec} shows noise spectra for a magnetically quiet day (upper plot) and using magnetometer data one day after a thunderstorm passed the site (lower plot).
All spectra lie above the ET design model because mitigation is not considered. Noise correlations are represented by the ET1-ET2 cross-PSD. Correlated noise is also the reason why the cross-PSD between ET1 and the null stream (in this curve including the multiplication with $\sqrt{3}$) does not match the ET1 PSD (as should be the case in the absence of correlated noise \cite{GNH2022}). The null-stream PSD tends to be lower than the PSD of individual ET-LF interferometers as we would expect from negative noise correlations. 

\begin{figure}[ht!]
    \centering
    \includegraphics[width=\columnwidth]{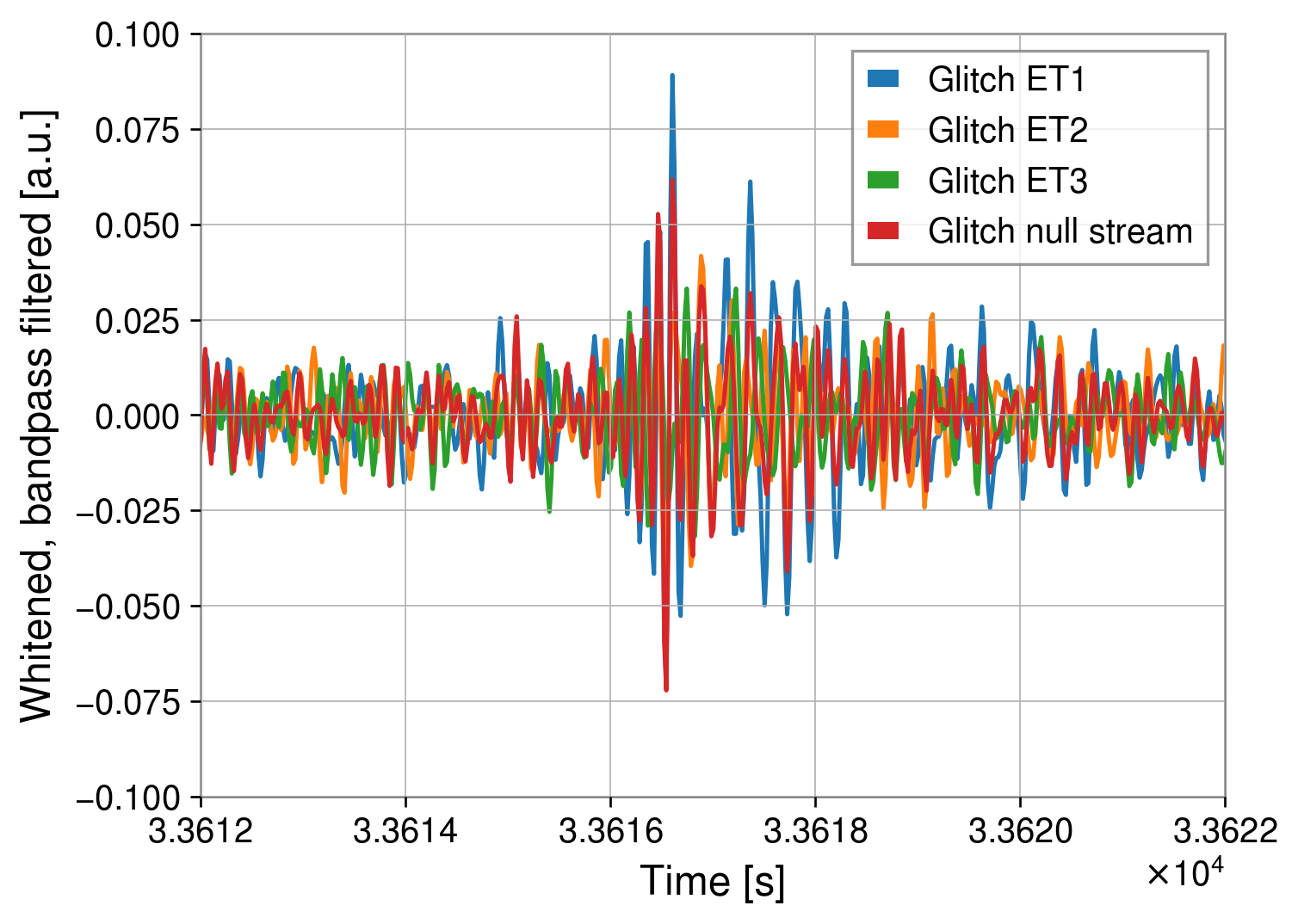}
    \includegraphics[width=\columnwidth]{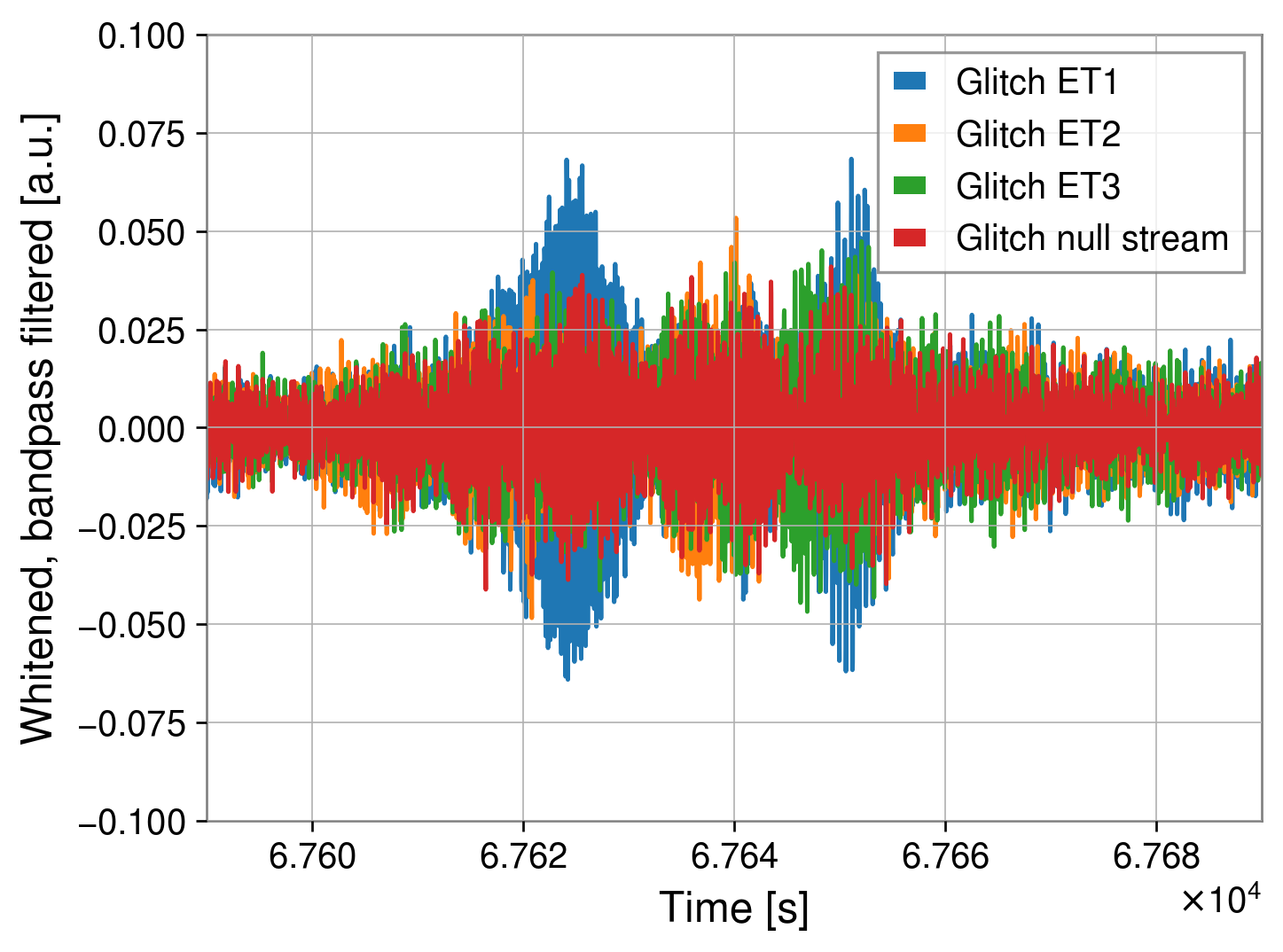}
    \caption{Glitch time series of the ET triangle. The upper time series spans 10\,s, the lower time series spans 100\,s. The glitches originate from transients observed with the Terziet borehole seismometer.}
    \label{fig:glitchtime}
\end{figure}

Finally, we look at the case of glitches. The point to check is if the null stream provides a model of the environmental glitches that would help to subtract them from the data of the individual interferometers. In figure \ref{fig:glitchtime}, we show two examples of glitches identified by the Omicron pipeline. Here we do not include magnetic noise in the simulation, and so, both glitches are produced by the seismic field through gravitational coupling. The time series were filtered to produce the plot: first the data are whitened using a full day of data, and then filtered with a 2\,Hz -- 10\,Hz band-pass filter. The segment of the time series in the upper plot is 10\,s long and in the lower plot 100\,s long. 

As intended, the glitches do not appear with the same shape in the three ET-LF interferometers. The physical explanation for this would be that the two test masses closest to each other from two different interferometers have a distance of 400\,m in our simulation, and a seismic disturbance needs to travel from one to the other test mass changing the waveform on its way. Note that the effect of propagation on the waveform is not entirely trivial because a seismic disturbance is generally a mix of polarizations each traveling with a different speed. Similarly, magnetic glitches would appear with different waveforms in two interferometers because of differences in how magnetic-field lines are distorted by the present vacuum infrastructure around each test mass. 

The glitches in the null stream do not resemble the glitches in the individual interferometers, while for glitch mitigation, the shape of the glitch in the null stream must accurately match the glitch in the individual interferometers \cite{GNH2022,Narola2025}. This would be the case if the glitch only appeared in one interferometer, but the coincidence of glitches in two or three interferometers means that null stream cannot provide faithful information. In the extreme case where exactly the same force acts on the pairs of closest test masses, the glitch would have strongly suppressed amplitude in the null stream due to the relative minus sign in the interferometer responses (see eq.~\eqref{eq:cohifo}). This extreme case is more likely to happen with Newtonian glitches when the corresponding seismic disturbance travels perpendicularly to the direction of the arm.
Consequently, the null stream is of no use to model the Newtonian and magnetic glitches in the individual interferometers. 

\section{Conclusions}
In this paper, we have presented a framework for time-domain simulations of ET data in its $\Delta$-configuration. In addition to the Gaussian noise model calculated from the ET sensitivity curve, environmental noise from magnetic and gravitational fluctuations at the test masses is included. The model implements noise correlations between ET-LF interferometers, which are needed to simulate sensitivity limitations of stochastic searches, the noise spectrum of the null stream, and which automatically produces a consistent simulation of noise transients stemming from the environmental fields. 

The two main points that had to be investigated are whether the null stream can be used to model the noise spectrum of individual detectors, and whether the null stream can be used to model glitches in the individual interferometers. Both could help to improve stochastic and transient GW searches. 

First, we have shown that the magnetic and Newtonian gravitational fields will produce a large number (of order a million glitches per year) of coincident glitches in all three ET-LF interferometers. Second, we find that environmental-noise correlations pose a strong sensitivity limitation to stochastic GW searches (consistent with previous work), and that they also have an important impact on the spectrum of the null stream. Third, we find that the waveform of magnetic and Newtonian glitches in the null stream does not match the waveform of glitches in the individual interferometers. These results show that the null stream is ineffective to model noise spectra and magnetic and Newtonian glitches in the ET triangle for noise mitigation. 

This does not exclude the possibility that the null stream can be used to mitigate some glitches. For example, faulty subsystems have led to glitches in the LVK detectors, which could also happen in ET. Since these glitches are independently produced in one of the interferometers, they would appear with identical shape (but with $\sqrt{3}$ lower SNR) in the null stream. Also some of the environmental glitches might not appear coincidentally in ET interferometers. For example, glitches produced by stray light have a complex nonlinear coupling mechanism, and it is conceivable that the seismic field produces some of them in one interferometer without causing a stray-light glitch in another interferometer. The ET-LF design addresses stray-light noise with an extensive baffle system, improved active seismic isolation and inter-platform sensing and control, but maybe this will not fully solve the stray-light problem. 

In conclusion, the ET triangle configuration can be expected to be subject to the same kind of problems with environmental noise as was observed in the two Hanford interferometers of initial LIGO. Not only is the null stream ineffective with the modeling and therefore mitigation of noise, our results indicate that the ET triangle requires orders-of-magnitude mitigation of Newtonian and magnetic noises to enable the full ET observational capabilities. 

\acknowledgments
We are grateful to Tito Dal Canton, Thomas Dent, Boris Goncharov, Michele Maggiore, Alex Nitz, Jacopo Tissino for providing information about the impact of glitches on GW data analysis and for comments on the manuscript. This work received support under the FIS-2024-01924 Advanced Grant \emph{Deep Loop Shaping for Gravitational-wave Detection}.

\bibliographystyle{apsrev}
\bibliography{references}

\end{document}